\documentclass[11pt]{amsart}
\usepackage{amsaddr}
\usepackage{amssymb}
\usepackage{enumerate}
\usepackage{enumitem}
\usepackage{amsmath,amsthm,marvosym,wasysym,mathdots}
\usepackage{bbm}
\usepackage{natbib}
\usepackage{color}
\usepackage{setspace}
\usepackage{epsfig}
\usepackage{hyperref}
\usepackage{subcaption}
\usepackage{pbox}
\usepackage[stable]{footmisc}
\hypersetup{%
  colorlinks=false,% hyperlinks will be black,
  pdfborder = {0 0 0.5 [3 0]}
}
\usepackage{breakurl}
\newcommand{\indic}[1]{1\hspace{-2.1mm}{1}_{\{#1\}}} %Indicator Function
\usepackage{caption}
\usepackage{multirow}
\usepackage[left=1in,right=1in,top=1.1in,bottom=1in]{geometry}

\newtheorem{theorem}{Theorem}
\newtheorem*{theorem*}{Theorem}

\newtheorem{corollary}[theorem]{Corollary}

\newtheorem{definition}[theorem]{Definition}

\newtheorem{lemma}[theorem]{Lemma}

\newtheorem{proposition}[theorem]{Proposition}

\theoremstyle{remark}
\newtheorem*{rem}{Remark}

\def\N{{\mathbb N}} % positive integers
\def\Q{{\mathbb Q}} % rationals
\def\R{{\mathbb R}} % reals
\def\P{{\mathbb P}} % probability
\newcommand{\EE}{{\mathord{I\kern -.33em E}}}
\def\E{{\mathbb E}} % expectation
\def\1{1{\hskip -3.3 pt}\hbox{I}}

\providecommand{\varitem}{}


\numberwithin{equation}{section}
\numberwithin{theorem}{section}

\usepackage{float}
\usepackage{hhline}
\usepackage[font=small,skip=0pt]{caption}
\usepackage{bm}
\usepackage{pbox}
\numberwithin{equation}{section}
\numberwithin{theorem}{section}

\usepackage{algorithm} 
\usepackage{algpseudocode} 
\algdef{SE}[DOWHILE]{Do}{doWhile}{\algorithmicdo}[1]{\algorithmicwhile\ #1}%
\algblock{Inputs}{EndInput}
\algnotext{EndInput}
\algblock{Output}{EndOutput}
\algnotext{EndOutput}

\begin{document}

\title{Acceptable Bilateral Gamma Parameters}
\author{Yoshihiro Shirai}
\date{\today} 
\email{yshirai@umd.edu}
\address{Department of Mathematics, University of Maryland, College Park}
 
\begin{abstract}
The purpose of this paper is to utilize statistical methodologies to infer from market prices of assets and their derivatives the magnitude of the set of a measure $\mathcal{M}$ that defines acceptance sets of risky future cash flows. We assume that $\mathcal{M}$ contains the collection of bilateral gamma random variables, and estimate upper and lower boundaries of the compensation needed for a given bilateral gamma distributed future cash flow to be acceptable. We show that prospects theory provides a natural interpretation of the behaviors implied by such boundaries, which are not compatible with expected utility theory. Boundaries for bilateral gamma risk neutral scale parameters for given speed parameters are also estimated and tested against market data and, in particular, comparisons are made with known empirical facts about the magnitude of the acceptance set of a common class of risk measures.
\end{abstract}

\maketitle

\section{Introduction}\label{Intro}
The definition of \textit{acceptable} risks, based on the axiomatization of the concept of coherent risk measure given in \cite{Artzner1999} and their convex generalization (\cite{FollmerSchied}), is a major recent advance in mathematical finance, as, among other applications, it provides an operative framework for superhedging in incomplete markets. Starting from a monetary measure, such as Value at Risk, that only satisfies the basic requirements of monotonicity and cash invariance, practical considerations (e.g. that the combined exposure of two trading desks ought to be less risky than that of the two desks taken separately, or that lack of liquidity may affect the future net worth of a single, large, position) lead one to require that a measure of risk also satisfy subadditivity and positive homogeneity. A measure of risk $\rho$ then defines a set $\mathcal{A}^{\rho}$ of acceptable risks as those random variables $X$ such that $\rho(X)\geq 0$. Conversely, it is possible to show that given a cone $\mathcal{A}$ of acceptable risks, the functional
\begin{align}\label{rhoA}
\rho(X)=\inf\{m\in\R: m+X\in \mathcal{A}\}
\end{align}
satisfies monotonicity, cash invariance, subadditivity and positive homogeneity. Based on convex duality, a risk measure is also specified by a set of equivalent probability measures $\mathcal{M}$ as
\begin{align}\label{rhoM}
\rho(X) = \inf_{\Q\in\mathcal{M}}\E^{\Q}[X].
\end{align}
The class $\mathcal{M}$ can be interpreted as the set of possible and credible macroeconomic/financial models, so that \ref{rhoM} is referred to as the robust representation of $\rho$, and risk measures become natural tools for the purpose of modeling uncertainty. For convex risk measures, a penalty $\alpha(\Q)$ is added to \ref{rhoM} to take into account that some models $\Q\in\mathcal{M}$ may be more or less plausible than others. 

Typical examples of risk measures are those based on certainty equivalent, such as the entropic risk measure, which are known in general as \textit{utility-based shortfall risk measures} and are defined by the acceptance set
\begin{align*}
\mathcal{A}=\{X:\E[u(X)]\geq u(c)\}
\end{align*}
for a given convex utility $u$ and a threshold $c$, and those obtained by modifying the tails of the underline statistical measure $\P$, such as the expected shortfall, which are known in general as \textit{spectral risk measures} and are defined by the Choquet integral
\begin{align*}
\rho(X)=\int_0^{\infty}\Psi(\P(X^+\geq a))da-\int_0^{\infty}\hat{\Psi}(\P(X^-\geq a))da,
\end{align*}
where $\Psi:[0,1]\rightarrow [0,1]$ is increasing and convex and $\hat{\Psi}(u)=1-\Psi(1-u)$. 

As the above examples confirm, relatively little is known in general about the set $\mathcal{M}$. Note, however, that a risk measure is an expected value under a worst case scenario measure, and, as such, it defines a minimal current valuation (or maximal bid price) of the future cash flow $X$, while $-\rho(-X)$ gives a maximal valuation (or the minimal ask price). Assuming that market prices of traded assets are random variables whose distribution belong to a specific class and is determined by a set $\Theta \in\R^D$ of parameters, observed market prices imply specific boundaries for the set $\Theta$ and, in turn, for $\mathcal{M}$. For instance, if $\mathcal{M}$ is (or contains) the class of normal random variables parameterized by pairs $(\mu,\sigma^2)$ of mean and variance of assets returns, one can ask what are maximal and minimal bounds for $\mu$ given $\sigma^2$ that are implied by historically observed pairs $(\mu,\sigma^2)$ of traded assets, in turn estimated from market prices. These bounds are then naturally interpreted as structural limits for the reward $\mu$ given the risk $\sigma^2$ that the economic system can offer without compromising its financial stability, as defined by the regulator. 

To fix a reference framework, consider a market composed only of one risky asset with log return $X$ and a riskless one in zero net supply with zero risk free rate. Then,
\begin{align}\label{EqPr}
1=\E^{\Q}[e^X]=\E[\eta e^X],
\end{align}
where $\Q$ is a risk neutral measure, $\eta$ the corresponding stochastic discount factor. If the distribution of $X$ under the statistical measure $\P$ is parameterized by $\theta\in\R^D$, and assuming the existence of a representative investor with utility $U$ defined by a set of parameters $\xi\in\R^m$, there is a function $V:\R^D\times\R^m\rightarrow \R$ that evaluates to $1$ at $(\theta,\xi)$. Specifically (see e.g. \cite{MadanEntropy}), the risk neutral density (with respect to the log return) is given by
\begin{align}\label{thetaxi}
h(x,\theta,\xi)=\frac{U_{\xi}'(e^x)f_{\theta}(x)}{\int_{\R}U_{\xi}'(e^s)f_{\theta}(s)ds},
\end{align}
where $f_{\theta}$ is the statistical density of $X$. Based on \ref{EqPr} and \ref{thetaxi}, if the prospects offered by the risky asset suddenly deteriorate, 1with $\theta$ replaced by a riskier $\theta'$, a decrease in the equilibrium risk free rate is needed to compensate. In extreme cases, however, investors may no longer be allowed to hold such an asset which will be liquidated and may, ultimately, stop trading in some markets. As an example, one may think of pension funds, which are not allowed to hold speculative grade bonds, or to those asset classes, such as hedge funds, that are only reserved to institutional investors.

In the case of normal returns, as it is well known (\cite{Markovitz1952}, \cite{Tobin1958}, \cite{Sharpe1964}, \cite{Lintner1965}), the efficient frontier essentially provides the upper limit for the reward $\mu_p$ given a risk defined by $\sigma^2$, and also the lower one, as this is the upper limit for a short position. In general, however, this result lies on the assumption that investors have mean-variance preferences, and that, in particular, they are expected utility maximizers. Empirical observations, on the other hand, have shown in many occasions that asset returns are not compatible with such axioms - a well known example being the equity premium puzzle, according to which U.S. equity risk premia over Treasury Bills rates reflect an implausible level of aversion to risk under expected utility theory (\cite{MehraPrescott1985}).

An alternative to expected utility theory, termed ``prospects theory'', is based on a series of experiments conducted by psychologists D. Kahneman and A. Tverski (\cite{KahnemanTverski1979}). One of their results, in particular, is that humans tend to be risk seekers rather than risk averse in the case of pure losses prospects. For instance, the prospect of winning $1000$ dollars with probability $1/2$ and winning zero otherwise is generally dominated by the prospect of winning $500$ dollars with probability $1$, but the prospect of losing $1000$ dollars with probability $1/2$ and losing zero otherwise dominates the prospect of losing $500$ dollars with probability $1$, independently of initial wealth. Based on such evidence, one is then led to interpret an asset's return as the sum of two prospects, one consisting of pure gains, and the other one of pure losses, and investors rank different assets' returns based on the expectations and variances $(\mu_p,\sigma_p^2,\mu_n,\sigma_n^2)$ of gains and losses. In particular, higher variance of losses is compensated, ceteris paribus, by lower expectation $\mu_p$ of the gains. 

The bilateral gamma distribution (\cite{KuchlerTappe}) and its multivariate version (\cite{MadanMBG}) provide a natural modeling framework for such a preference specification for several reasons. Firstly, it is the difference of two independent gamma variates, interpretable as gains and losses, and it is completely specified by the vector $(\mu_p,\sigma_p,\mu_n,\sigma_n)$ of their expected values and standard deviations. Secondly, even in a continuous time setting, the bilateral gamma process is the difference of two independent gamma processes, while, for instance, path realizations of diffusion processes have infinite variation. Thirdly, the bilateral gamma distribution provides a very good fit to the (log) returns distribution implied by time series of returns and also by options prices (\cite{KuchlerTappe}), which shows that it is more suitable than, e.g., the normal distribution for the purpose of modeling asset returns. Finally, as shown below, the expected utility of an asset with bilateral gamma return $X$ is a function $F:(\mu_p,\sigma_p,\mu_n,\sigma_n)\rightarrow \E[u(X)]$, increasing in $\mu_p$, and decreasing in $\sigma_p$, $\mu_n$ and $\sigma_n$, so that under expected utility theory variations in $(\sigma_p,\mu_n,\sigma_n)$ are compensated by variations of equal sign in $\mu_p$. 

Based on this considerations, we assume in this paper that the set of credible models $\mathcal{M}$ includes the set of bilateral gamma random variables, and we learn bounds $f_M,f_m:(\sigma_p,\mu_p,\sigma_n)\rightarrow \mu_p$ for $\mu_p$ given risks $(\sigma_p^2,\mu_n,\sigma_n^2)$ via quantile and/or distorted linear and/or Gaussian process regression. 

An interesting result obtained is that both boundaries are generally increasing in $(\sigma_p,\mu_p)$, but decreasing in $\sigma_n$, suggesting that investors, independently of their wealth, seek for lower (resp. higher) risk when it comes to purely positive (resp. negative) processes. We test the boundaries computed by assessing how well their implied performance measures (Sharpe ratio and acceptability index) compare with those typically observed in the financial markets. Furthermore, we investigate the linearity of $f_M$ and $f_m$ by comparing the results of a linear lower dimensional embedding and a nonlinear one, and we show through a simple variation of a Lucas tree economy \cite{Lucas1978} that the behaviors observed are indeed consistent with prospects theory.

%Our analysis differs from the one performed in \cite{MadanEntropy} as there a lower valuation (bid price) $V$ of a dollar invested in the asset is regressed against gains and losses moments, while here $\mu_p$ is directly regressed against the upside volatility and downside risks. 

%With respect to \cite{MadanAnchored}, the dataset here analyzed is more ample, a theoretical interpretation of the risk seeking behaviors is proposed, and possible nonlinear embeddings are investigates.

Finally, we move our attention to the risk neutral world, based on the suggestive interpretation given in \cite{MadanEntropy} that, for bilateral gamma returns, the scale parameters $(b_p,b_n)$ determine the structure of limit orders, while the speed parameters $(c_p,c_n)$ determine that of market orders. It is then natural to assume that a relationship exists between the two pairs of parameters, in the sense that for given $(c_p,c_n)$, the scale parameters $(b_p,b_n)$ are bounded to a specific range, as the structure of market orders cannot be too independent from that of limit orders and viceversa. As done for the statistical moments, the boundaries of such range are learned through quantile and distorted regression. In this case, we determine theoretical boundaries as well based on the well known robust representation of spectral risk measures (\cite{MDV}), and evidence is offered of their comparability with the empirically estimated ones. 

The rest of the paper is organized as follows. First we show that for bilateral gamma returns, risks and compensations are identified by the vector $(\sigma_p,\mu_n,\sigma_n)$ and $\mu_p$ respectively. Empirical observations are reported in section 3, and the variation on Lucas Tree model is presented in section 4. Risk neutral parameters are analyzed in 5. Section 6 concludes.
%
%The rest of the paper is organized as follows. The bilateral gamma process and dimensionality reduction techniques are described in section \ref{BG} and\ref{DimRed}, while manifold discovery results for our database are reported in section \ref{BGDim}. Boundaries of the low dimensional embedding is characterized via gaussian process regression in section \ref{Parametrization}.

\section{Bilateral Gamma Returns}\label{BG}  
\subsection{From Brownian Motion to Bilateral Gamma Process}
Given its central role in this paper, the construction and properties of the bilateral gamma process are reviewed in this section. In \cite{BlackScholes}, F. Black and M. Scholes proposed to model the dynamics of log-returns as a Brownian motion (GBM), as prices exhibit exponential growth and on the assumption, rooted in an entropy maximization argument (\cite{MadanEntropy}), that log returns are asymptotically normally distributed. 

However, returns exhibit heavier tails than those implied by the normal distribution (\cite{Fama1965}) and frequent discontinuities in their path trajectories. In addition, risk aversion results in periods of intense trading, determined by widespread selling in securities, alternating with lower activity ones, thus implying that returns' quadratic variation is not linear in time. It also results in higher demand for out of the money (OTM) than for the corresponding OTM calls, generating a volatility smile.
%\footnote{An interesting example to gain intuition is as follows. The actual cost incurred by implementing strategies that dynamically replicate equity option prices for, say, a deep out of the money (OTM) put is higher, on average, than the cost implied by the Black-Scholes formula for the at the money (ATM) implied volatility $\sigma_{imp}$ (and other inputs being held constant). In other words, the realized volatility of the portfolio replicating a deep out of the money option tends to be higher than $\sigma_{imp}$, or, which is the same, the volatility realized by the replicating strategy reaches $\sigma_{imp}$ in a time shorter than $T$. A natural interpretation of this is the assertion that economic activity runs at a pace different (faster, on average) than actual calendar time.}

Another entropy maximization argument then suggests modeling economic time as a gamma process, and stock market log returns as Brownian motion evaluated at such gamma time. The resulting process, pioneered by D. Madan and E. Seneta (\cite{MadanSeneta}) and termed the variance gamma process, is a pure jump Levy process with infinite activity and finite variation. In fact, such process is the difference of two i.i.d. gamma processes, which naturally correspond to gains and losses. Finally, motivated by the fact that downward jumps in prices are generally higher than upward ones, the bilateral gamma process is defined as the difference of two independent gamma processes with different shape and scale parameters (\cite{KuchlerTappe}). The gains and losses increments have BG distribution $\beta\Gamma(b_p,c_p,b_n,c_n)$, defined by the convolution
\begin{align*}
\beta\Gamma(b_p,c_p,b_n,c_n)=\Gamma(b_p,c_p)\ast \Gamma(-b_n,c_n),
\end{align*}
where $b_p,c_p,b_n,c_n>0$ and, for $\alpha>0$, $\lambda\in\R$, a $\Gamma(\lambda,\alpha)$-distributed random variable has density
\begin{align*}
f(x)
=\frac{1}{\Gamma(\alpha)|\lambda|^{\alpha}}|x|^{\alpha-1}
	e^{-|x|/|\lambda|}\left(\indic{\lambda>0}(x)\indic{x>0}(x)
	+\indic{\lambda<0}(x)\indic{x<0}(x)\right), \ x\in\R
\end{align*} 
with $\Gamma(\alpha)$ the Gamma function at $\alpha$. Then, expected value and standard deviation of gains and losses, denoted respectively by $\mu_p$, $\sigma_p$, $\mu_n$ and $\sigma_n$, are given by
\begin{align*}
\mu_p=c_pb_p, \ \sigma_p=\sqrt{c_p}b_p, \ \mu_n=c_nb_n, \ \sigma_n=\sqrt{c_n}b_n.
\end{align*}
By the convolution theorem, the characteristic function of the increments in $t$ units of time is
\begin{align}\label{BGphi}
\varphi_t(u)
=\left(1-iub_p\right)^{-tc_p}
\left(1+iub_n\right)^{-tc_n},
\end{align}
and it follows easily from \ref{BGphi} that BG densities are stable under convolution and are infinitely divisible, and so the BG process is a well defined Levy process. From formula \ref{BGphi} and the Levy-Khintchine representation we also deduce its Levy density to be
\begin{align*}
k(x)=\left(
	\frac{c_p}{x}e^{-x/b_p}\indic{(0,\infty)}(x)
	+ \frac{c_n}{|x|}e^{-|x|/b_n}\indic{(-\infty,0)}(x)
	\right), \ x\in\R
\end{align*}
which shows that a BG process enjoys the self decomposability property.\footnote{A random variable $X$ is self decomposable if for any $0<c<1$ there is an independent random variable $X_C$ such that $X\,{\buildrel d \over =}\ cX+X_c$. A Levy process enjoys the self decomposability property if its increments are self decomposable.} Then (see \cite{CGMY} and the references therein) a BG distributed random variable $X$ is a limit law, i.e. there are centering and scaling constants $\{c_n\}_{n\in\N}$ and $\{b_n\}_{n\in\N}$ and a sequence $\{Z_k\}_{k\in\N}$ of i.i.d. random variables such that the distribution of $b_nS_n+c_n$ converges in distribution to $X$, where $S_n=\sum_{k=1}^nZ_k$. This is a remarkable property, since if returns consist of some average of a large number of independent news or other type of influences, it is reasonable to expect that their distribution should be well approximated by a limit law. In the GBM case such law is the Gaussian, but, as noticed in \cite{CGMY}, there is ``no compelling economic motivation" for the scaling constants to be $\sqrt{n}$ as in the classical central limit theorem. 

Evidence of the goodness of fit of the BG density to returns distributions is presented in \cite{KuchlerTappe}, where, using data on DAX between 1996 and 1998, it is shown that the null hypothesis that the log returns distribution is in the BG class is not rejected. Furthermore, as proved in \cite{KuchlerTappe}, for all BG parameters there exists a measure $\Q$ equivalent to $\P$ such that, under $\Q$, the discounted exponential BG process is a (local) martingale and an exponential BG process, and one typically succeed in fitting the option prices surface, at least for a single fixed maturity, through an exponential BG process.

\subsection{Bilateral Gamma Returns under Expected Utility Theory}\label{EUT}

The notion and characterizations of second order stochastic dominance (SSD) are recalled below (see \cite{RothschildStiglitz}).
\begin{definition}
Given random variables $X$ and $Y$, one says that $X$ first (resp. second) order stochastically dominates $Y$, i.e. $X\succeq_1 Y$ (resp. $X\succeq_2 Y$) if and only if $\E[u(X)]\geq \E[u(Y)]$ for every increasing (resp. increasing and concave) real valued function $u$. 
\end{definition}
\begin{theorem}\label{FSDTh}
Let $X$ and $Y$ be random variables with distribution functions $F$ and $G$ respectively. Then, $X\succeq_1 Y$ if and only if $G(t)\geq F(t)$ for every $t\in\R$.
\end{theorem}
\begin{theorem}\label{SSDTh}
Let $X$ and $Y$ be random variables with distribution functions $F$ and $G$ respectively. Then, the following are equivalent
\begin{itemize}
[nolistsep,noitemsep]
\item[(i)] $X\succeq_2 Y$;
\item[(ii)] there are random variables $Z$ and $\varepsilon$ such that $Y\sim X+Z+\varepsilon$, $Z\leq 0$ and $\E[\varepsilon|X+Z]=0$;
\item[(iii)] $\int_{-\infty}^tG(s)ds\geq \int_{-\infty}^sF(s)ds$ for every $t\in\R$.
\end{itemize}
In addition, if $\E[X]=\E[Y]$, then the following are equivalent:
\begin{itemize}
[nolistsep,noitemsep]
\item[(i)] $X\succeq_2 Y$;
\item[(ii)] there is a random variable $\varepsilon$ such that $Y\sim X+\varepsilon$ and $\E[\varepsilon|X+Z]=0$;
\item[(iii)] $\E[u(X)]\geq \E[u(Y)]$ for every $u$ concave.
\end{itemize}
\end{theorem}
\begin{corollary}\label{SSDVar}
Suppose $X\succeq_2 Y$. Then, $\E[X]\geq \E[Y]$ and if $\E[X]=\E[Y]$ then $V(X)\leq V(Y)$.
\end{corollary}
\begin{proof}
That $\E[X]\geq \E[Y]$ if $X\succeq_2 Y$ follows immediately from the fact that the identity is non decreasing and concave. If $\E[X]=\E[Y]$, then $\E[u(X)]\geq \E[u(Y)]$ for every $u$ concave, and so, setting $u(x)=-x^2+\E[X]$, one obtains $V(X)=\E[X^2-\E[X]]\leq \E[Y^2-\E[X]]=V[Y]$.
\end{proof}
Thus, for bilateral gamma returns, SSD implies higher expected gains and/or lower expected losses, and, for equal expected gains and losses, lower standard deviation of gains and/or losses. 
%When assets log returns have bilateral gamma distribution, the following proposition holds.
%\begin{proposition}
%Suppose $X\sim \beta G$ with gain and loss expectation and standard deviation $\mu_p^X$, $\sigma_p^X$, $\mu_n^X$, and $\sigma_n^X$. Then
%\begin{align*}
%\frac{d}{d\mu_p^X}\E[e^X]\geq 0, \
%	\ \frac{d}{d\mu_p^X}V[e^X]\leq 0, \\
%\frac{d}{d\sigma_p^X}\E[e^X]\leq 0, \
%	 \ \frac{d}{d\sigma_p^X}V[e^X]\geq 0, \\
%\frac{d}{d\mu_n^X}\E[e^X]\leq 0, \
%	\frac{d}{d\mu_n^X}V[e^X]\geq 0, \\
%\frac{d}{d\sigma_n^X}\E[e^X]\leq 0, \
%	\frac{d}{d\sigma_n^X}V[e^X]\geq 0.
%\end{align*}
%In particular, if $Y\sim \beta G$ with gain and loss expectation and standard deviation $\mu_p^Y$, $\sigma_p^Y$, $\mu_n^Y$, and $\sigma_n^Y$, then $e^Y\succeq e^Y$ implies at least one of the following inequalities: $\mu_p^X\geq \mu_p^Y$, $\sigma_p^Y\geq \sigma_p^X$, $\mu_n^X\geq \mu_p^Y$, $\sigma_n^Y\geq \sigma_n^X$.
%\end{proposition}
%\begin{proof}
%We have
%\begin{align*}
%\frac{d}{d\mu_p}\E[e^X]
%& = \frac{d}{d\mu_p}
%	e^{-\tfrac{\mu_p^2}{\sigma_p^2}
%		\log\left(1-\tfrac{\sigma_p^2}{\mu_p}\right)}
%	e^{-\tfrac{\mu_n^2}{\sigma_n^2}
%		\log\left(1+\tfrac{\sigma_n^2}{\mu_n}\right)}\\
%& = 
%	\left(1+\tfrac{\sigma_n^2}{\mu_n}\right)
%	^{-\tfrac{\mu_n^2}{\sigma_n^2}}
%\end{align*}
%\end{proof}
A partial converse of this statement is shown below, and is based on the following results.
\begin{theorem}\label{LR}
Let $X$ and $Y$ be random variables with densities $f$ and $g$. If the likelihood ratio $\tfrac{f}{g}$ is monotonically increasing, than $X\succeq_1 Y$. If the likelihood ratio is monotonically increasing on $(-\infty,x_0)\cup (x_1,\infty)$ and decreasing on $(x_0,x_1)$, with $x_0<x_1\in\R$, then $X\succeq_2 Y$.
\end{theorem}
\begin{proof}
See \cite{AliSSD} and the references therein.
\end{proof}
\begin{theorem}\label{AliSSDTh}
Let $X$ and $Y$ be two gamma distributed random variable with scale and shape parameters $(b,c)$ and $(b',c')$ respectively. Then,
\begin{itemize}
[noitemsep, nolistsep]
\item[(i)] if $b=b'$, then $c>c'$ iff $X\succeq_2 Y$;
\item[(ii)] if $c=c'$, then $b>b'$ iff $X\succeq_2 Y$;
\item[(ii)] $\frac{c}{c'}\leq \max(1,\tfrac{b'}{b})$ with strict inequality at least when $\tfrac{b'}{b}=1$ iff $X\succeq_2 Y$.
\end{itemize}
\end{theorem}
\begin{proof}
Based on showing that the assumptions of theorem \ref{LR} are satisfied. See \cite{AliSSD}.
\end{proof}
\begin{corollary}\label{SSDBG1}
Let $X$ and $Y$ be two gamma distributed random variables with scale and shape parameters $(b,c)$ and $(b',c')$ respectively. Then, $X$ second order stochastically dominates $Y$ if $\E[X]\geq \E[Y]$ and $V[X]\leq V[Y]$ with at least one strict inequality. Similarly, $-X$ second order stochastically dominates $-Y$ if $\E[X]\leq \E[Y]$ and $V[X]\leq V[Y]$ with at least one strict inequality.
\end{corollary}
\begin{proof}
Suppose $\E[X]\geq\E[Y]$ and $V[X]\leq V[Y]$ with at least one strict inequality, i.e. $bc\geq b'c'$ and $\ b^2c\leq b'^2c'$ with at least one strict inequality. Then, $\tfrac{c}{c'}\geq \tfrac{b'}{b}$, and 
\begin{align*}
\frac{b'}{b}
=\frac{b'^2c'}{b^2c}\frac{bc}{b'c'}>1
\end{align*}
so $X\succeq_2 Y$ by Theorem \ref{AliSSDTh}. The result for $-X$ and $-Y$ follows from adapting Theorem \ref{AliSSDTh} to the case of the negative of gamma distributions.
\end{proof}
\begin{corollary} \label{SSDBG2}
Let $X^+$, $X^-$, $Y^+$, $Y^-$ be four gamma distributed random variable with scale and shape parameters $(b_p,c_p)$, $(b_n,c_n)$, $(b'_p,c'_p)$ and $(b'_n,c'_n)$ respectively. Then, $X:=X^+-X^-$ second order stochastically dominates $Y:=Y^+-Y^-$ if $\E[X^+]\geq \E[Y^+]$, $\E[X^-]\leq \E[Y^-]$, $V[X^+]\leq V[Y^+]$, and $V[X^-]\leq V[Y^-]$ with exactly one strict inequality.
\end{corollary}
\begin{proof}
Suppose for instance $\E[X^+]>\E[Y^+]$. Then, by Corollary \ref{SSDBG1}, $X^+\succeq_2 Y^+$, and so, for all $t\in [0,\infty)$
\begin{align*}
\int_0^tF^+(s)-G^+(s)ds\leq 0,
\end{align*}
where $F^+$ and $G^+$ denote the cumulative distribution function of $X^+$ and $Y^+$ respectively. Then, using Tonelli's theorem,
\begin{align*}
\int_0^tF(s)-G(s)ds
=\int_0^{\infty}\int_0^tF^+(s-\xi)-G^+(s-\xi)dsdF^-(\xi)
\leq 0,
\end{align*}
where $F^-$ is the (common) distribution of $-X^-$ and $-Y^-$, and the conclusion follows from Theorem \ref{SSDTh}. The other cases are similar.
\end{proof}
Based on the last corollary and transitivity of SSD, the observation that a positive variation in $\mu_p$ can compensate a positive variation in any among the upside volatility $\sigma_p$, the expected loss prospect $\mu_n$ or the downside volatility $\sigma_n$ is evidence of investors' risk seeking behaviors. 

Note that $\mu_p$ is not, in general, a ``reward'' accessible to an investor holding the asset. In fact, for a given time horizon $T$ the expected return for holding the asset is the value $\mu(T)$ that satisfies $S_0e^{\mu(T)}=\E[S_0e^{X_T}]$, and so the variation
\begin{align}\label{Reward}
\lim_{T\downarrow 0}\frac{\mu(T)}{T}=\int_{\R}(e^x-1)k(x)dx 
= (1-b_p)^{-c_p}(1+b_n)^{-c_n}
\end{align}
better serves this purpose. Thus, we refer to $\mu_p$ as a ``compensation'' for the risks $(\sigma_p,\mu_n,\sigma_n)$.

\subsubsection{Log-Returns and Kelly's Criterion}

In the case log returns are assumed to be bilateral gamma variates, these results cannot hold anymore, since, for instance, an increase in $\sigma_p$ and/or $\sigma_n$ implies higher expected value of the return, and it cannot imply second order stochastic dominance. However, a traditional assumption in the financial and economics literature, justified by some evidence (\cite{Arrow1971}), is to assume that investors maximize log-returns. In our context, such an assumption implies that an asset is preferred to another one if and only if the expected log-return is higher. More generally, for asset allocation problems, logarithmic utility yields the best return in the long run, assuming the investor faces a long sequence of investment decisions (\cite{Kelly1956}, \cite{Merton69}, \cite{Cover1991}), but for an investor with a short/medium term horizon, a logarithmic utility will not capture aversion to short term high volatility (\cite{Samuelson1979}), thus leading to consider a utility specification with a coefficient of relative risk aversion (CRRA) bounded below by 1.\footnote{In fact, several empirical studies provide evidence for this to be the case (see e.g. \cite{FriendBlume}).} It then follows from the results of this section and proposition \ref{ULog} below that, for a reasonable utility specification such as $u(\log(\cdot))$, risks and their compensation are captured by $(\sigma_p,\mu_n,\sigma_n)$ and $\mu_p$ respectively even when log-returns belong to the bilateral gamma class.
\begin{proposition} A strictly increasing and concave function $v\in C^2\left((0,\infty)\right)$ has CRRA coefficient greater than $1$ if and only if there is a strictly increasing and concave function $u\in C^2(\R)$ such that $v(x)=u(\log(x))$ for every $x\in (0,\infty)$.
\end{proposition}\label{ULog}
\begin{proof} Suppose such a $u$ exists. Then, for all $x\in (0,\infty)$, $u''(\log(x))\geq 0$ and $u'(\log(x))< 0$
\begin{align*}
x\frac{v''(x)}{v'(x)}=-x\frac{\tfrac{d^2}{dx^2}u(\log(x))}{\tfrac{d}{dx}u(\log(x))}
=1-\frac{u''(\log(x))}{u'(\log(x))}\geq 1.
\end{align*}
On the other hand, if $v$ has CRRA bounded below by $1$, then, setting $u(y)=v(e^y)$ for every $y\in\R$, we obtain $u'(y)=v'(e^y)e^y>0$ and $u''(y)=v''(e^y)e^{2y}+v'(e^y)e^y\leq 0$.
\end{proof}
%We also observe that there is a one to one relationship between the moments $(\mu_p,\sigma_p,\mu_n,\sigma_n)$ and the shape and scale parameters $(b_p,c_p,b_n,c_n)$ of a bilateral gamma distributed random variable. Thus traded risks-compensation lie on a four dimensional manifold, unless investors only care of some lower dimensional subset of  to expect a lower dimensional embedding of such manifold. 

\section{The Acceptance Set}\label{DimRed}

\subsection{Learning the Boundaries}
%While assessing portfolios of assets that are on the mean-variance efficient frontier, an inverse relationship between risk (variance) and expected return is established. On the other hand, there is no immediate justification for there to exists a similar relationship when a single asset is being assessed, so that the range of rewards/compensations for given risks may be large, and there may be no lower dimensional embedding for the risks-compensation manifold itself.
%However, it is observed in \cite{MadanAnchored} that the supply of derivatives such as deep OTM put and call options, together with the market clearing condition, narrows the variability of compensation given risks. For instance, if the compensation for an asset is low compared to its risks, short positions  of a large increase in upside risk $\sigma_p$ not compensated by the compensation $\mu_p$, would create an excess of demand in the corresponding OTM call option market, and similarly for $\mu_n$ and $\sigma_n$. 

As mentioned in the introduction, not all quadruples $(\mu_p,\sigma_p,\mu_n,\sigma_n)$ can be traded, or, in other words, there are structural limits to how high and/or low is the level of rewards that can be offered for given risks. In order to determine such limits, moments of gains and losses were estimated for 184 stocks (whose ticker is reported in appendix A) for the period 01/01/2008 to 31/12/2020 using one year of data for each estimate.\footnote{Observations are results of likelihood optimization, so $1\%$ of outliers were excluded.}
% Figures \ref{Quantile1}-\ref{Quantile3} show scatter plots of the observations for each pair $(\mu_p,\sigma_p)$, $(\mu_p,\mu_n)$ and $(\mu_p,\sigma_n)$
Assuming the boundaries are defined by functions $f_m,f_M:(\sigma_p,\mu_n,\sigma_n)\rightarrow \mu_p$, we find $f_M$ and $f_m$ by solving, respectively,
\begin{align*}
& \min_{f\in \mathcal{F}}
	(1-\tau_M)\sum_i \left[\mu_p(i)-f_M(\sigma_p(i),\mu_n(i),\sigma_n(i))\right]^+
	-\tau_M\sum_i \left[\mu_p(i)-f_M(\sigma_p(i),\mu_n(i),\sigma_n(i))\right]^-,\\
& \min_{f\in \mathcal{F}}
	(1-\tau_m)\sum_i \left[\mu_p(i)-f_m(\sigma_p(i),\mu_n(i),\sigma_n(i))\right]^+
	-\tau_m\sum_i \left[\mu_p(i)-f_m(\sigma_p(i),\mu_n(i),\sigma_n(i))\right]^-,
\end{align*}
where $\mathcal{F}$ is a suitable class of functions which is here assumed to be the class of linear Gaussian process (GPR) regressors, $\tau_M=0.95$ and $\tau_m=0.05$. In our implementation of quantile GPR, the kernel hyperparameters were estimated using the standard loss function, while the regression coefficients are chosen to maximize the quantile loss function. Specifically, recall that GPR assumes 
\begin{align}\label{GPR}
\mu_p = \alpha + h(\sigma_p,\mu_n\sigma_n)^T\beta+f(\sigma_p,\mu_n\sigma_n)+\varepsilon,
\end{align}
where $\varepsilon$ is noise with variance $\sigma_{\varepsilon}^2$, $h$ is the map to features space (here assumed to be the identity), and where any finite number collection $\{f(\sigma_p,\mu_n\sigma_n)\}$ is assumed to have Gaussian distribution with mean $0$ and covariance function $\kappa((\sigma_p,\mu_n\sigma_n),(\sigma_p,\mu_n\sigma_n)')$. The prediction $\mu_p$ for $x=(\sigma_p,\mu_n,\sigma_n)$ given $n$ observations $(\mu_p^i,\sigma_p^i,\mu_n^i,\sigma^i_n)$ is then given by (see \cite{RasmussenWilliams})
\begin{align*}
\mu_p = \begin{bmatrix}
	\kappa(x^1,x) & ... &\kappa(x^n,x)
	\end{bmatrix}
	\left(\begin{bmatrix}
	(\kappa(x^1,x^1) & \dots & \kappa(x^1,x^n)\\
	 & \ddots & \\
	(\kappa(x^n,x^1) & \dots & \kappa(x^n,x^n)\\
	\end{bmatrix}_{i,j}+\sigma_{\varepsilon}^2I\right)^{-1}
	\begin{bmatrix}
	\mu_p^1\\
	\vdots\\
	\mu_p^n
	\end{bmatrix},
\end{align*}
where we let $x^i=(\sigma_p^i,\mu_n^i,\sigma_n^i)$. Here we take $\kappa$ to be the squared exponential kernel, with parameters estimated based on the standard loss function. The vector $\beta$ and the intercept $\alpha$  are instead chosen by minimization of the quantile loss function.

The linear estimates obtained for $f_m$ and $f_M$ are
\begin{align*}
f_m(\sigma_p,\mu_n,\sigma_n)=0.0017+0.2029\sigma_p+0.9951 \mu_n-0.3711\sigma_n, \\
f_M(\sigma_p,\mu_n,\sigma_n)=0.0017+0.2710\sigma_p+1.0102 \mu_n-0.2311\sigma_n.
\end{align*}
Note, in particular, the negative relationship between $\sigma_n$ and $\mu_p$. 

Similarly, for quantile GPR, $\tfrac{\partial f_m}{\partial \sigma_n}$ are always negative, while $\tfrac{\partial f_M}{\partial \sigma_n}$ are positive at all but two of 16 representative points (table \ref{table:QGPR}).

\begin{table}[h!]
\centering
 \begin{tabular}{| c | c | c | c | c | c |} 
 \hline $\tfrac{\partial f_M}{\partial \sigma_p}$
 		& $\tfrac{\partial f_M}{\partial \mu_n}$
 		& $\tfrac{\partial f_M}{\partial \sigma_n}$
 		& $\tfrac{\partial f_m}{\partial \sigma_p}$
 		& $\tfrac{\partial f_m}{\partial \mu_n}$
 		& $\tfrac{\partial f_m}{\partial \sigma_n}$\\
	\hline\hline
 	0.2667 &  2.4704 &  0.7577 & -0.0130 &  2.0042 & -0.2421\\
 	\hline
    0.8691 &  1.9402 & -1.3539 &  1.1140 &  1.8974 & -0.8361\\
 	\hline
    1.5243 &  1.9553 & -1.1134 &  1.4108 &  1.9274 & -1.2346\\
 	\hline
    1.0459 &  2.0254 & -0.4887 &  0.5666 &  1.9927 & -1.2635\\
 	\hline
    1.0867 &  1.9956 & -1.0836 &  0.8823 &  2.0199 & -1.2220\\
 	\hline
    0.4639 &  2.0065 & -1.4194 &  0.6053 &  2.0648 & -1.1568\\
 	\hline
    1.3013 &  2.0509 & -1.4681 &  1.2715 &  2.0128 & -1.4149\\
 	\hline
    0.9669 &  2.0019 & -0.2462 &  0.4477 &  1.9760 & -1.0806\\
 	\hline
    1.4434 &  2.2522 &  0.3978 &  0.5052 &  2.0026 & -0.5761\\
 	\hline
    0.9710 &  1.9465 & -0.9840 &  0.9472 &  1.8900 & -0.8995\\
 	\hline
    1.0653 &  1.9423 & -1.4702 &  1.3075 &  1.9230 & -0.9990\\
 	\hline
    0.9307 &  1.9594 & -0.5941 &  0.6044 &  1.9087 & -1.0416\\
 	\hline
    1.3390 &  2.0444 & -1.8664 &  1.4529 &  2.0287 & -1.5394\\
 	\hline
    0.8652 &  1.9872 & -1.3001 &  0.9281 &  2.0499 & -1.0898\\
 	\hline
    1.1957 &  2.0398 & -0.9586 &  0.9027 &  1.9967 & -1.3931\\
 	\hline
    0.9283 &  1.9956 & -0.0906 &  0.3913 &  1.9830 & -0.9539\\
 	\hline
\end{tabular}
\caption{Boundaries gradients (estimated via Quantile GPR), at 16 representative points.}
\label{table:QGPR}
\end{table}

Alternatively, $f_m$ and $f_M$ can be obtained via distorted least square (\cite{MDV}), i.e. solving
\begin{align}\label{DLF}
& \min_{f\in \mathcal{F}}
	\sum_i r_i^2\left(\Psi(q_i)-\Psi\left(q_i-\tfrac{1}{n}\right)\right),
\end{align}
where, for every $i$, 
\begin{align*}
r_i:=\mu_p(i)-f(\sigma_p(i),\mu_n(i),\sigma_n(i)
\end{align*}
is the residual corresponding to the $i$-th observation, $\Psi:[0,1]\rightarrow [0,1]$ is a distribution function (called a distortion) concave for $f_m$ and convex for $f_M$, and $q_i$ is the $i$-th quantile of the residual's empirical distribution.

The idea behind \ref{DLF} is as follows. First, $\Psi$ defines a distorted expectation $\mathcal{E}^{\Psi}[X]$ of a random variable $X$ with distribution function $F$, as the Stjielties integral with respect to the distribution function $\Psi\circ F$:
\begin{align*}
    \mathcal{E}^{\Psi}[X]
    : = \int_{\R}xd\Psi(F(x)).
\end{align*}
If $\Psi$ is concave, lower values of $X$ are weighted higher, thus implying $\mathcal{E}^{\Psi}[X]\leq \E[X]$, while the opposite is true if $\Psi$ is convex.
Next, given observations $\{x_i\}_{i=0}^n$ of $X$, $\mathcal{E}^{\Psi}[X]$ is estimated by
\begin{align*}
\sum_{i=1}^n x_{(i)}\left[ \Psi(F(x_{(i)}))-\Psi(F(x_{(i-1)}))\right]
\end{align*}
where $\{x_{i}\}_{i=1,...,n}$ is the ordered sample. If $F$ is unknown, this estimator can be replaced by
\begin{align*}
\sum_{i=1}^n x_{i}\left[ \Psi\left(q_i \right)-\Psi\left(q_i -\frac{1}{n}\right)\right],
\end{align*}
so the loss function \ref{DLF} corresponds to minimizing the estimated distorted expectation of the squared residual. By the tower property of (nonlinear) conditional expectations, the solution to problem \ref{DLF} minimizes the distorted squared distance to the estimate of $\mathcal{E}^{\Psi}[\mu_p|\sigma_p,\mu_n,\sigma_n]$ and so, for an appropriate distortion, it can be thought as a lower/upper bound to the range of compensation $\mu_p$ given the risks $(\sigma_p,\mu_n,\sigma_n)$. Following \cite{MDV}, we set $\gamma=0.75$ and define, for $u\in [0,1]$,
\begin{align}\label{MINMAXVAR}
\Psi(u)=1-\left(1-u^{\tfrac{1}{1+\gamma}}\right)^{1+\gamma}.
\end{align}
The linear estimates obtained for $f_m$ and $f_M$ obtained via distorted least square are
\begin{align*}
f_m(\sigma_p,\mu_n,\sigma_n)=0.0024+0.1118\sigma_p+0.9276 \mu_n-0.2596\sigma_n, \\
f_M(\sigma_p,\mu_n,\sigma_n)=0.0002+0.3604\sigma_p+1.0196 \mu_n-0.1798\sigma_n,
\end{align*}
while the gradient at 16 representative points of the GPR estimates is shown in table \ref{table:DLSGPR}.

\begin{table}[h!]
\centering
 \begin{tabular}{| c | c | c | c | c | c |} 
 \hline $\tfrac{\partial f_M}{\partial \sigma_p}$
 		& $\tfrac{\partial f_M}{\partial \mu_n}$
 		& $\tfrac{\partial f_M}{\partial \sigma_n}$
 		& $\tfrac{\partial f_m}{\partial \sigma_p}$
 		& $\tfrac{\partial f_m}{\partial \mu_n}$
 		& $\tfrac{\partial f_m}{\partial \sigma_n}$\\
	\hline\hline
 	-0.0887 & 1.9864 & 0.6572 & 0.0790 & 2.0073 & -0.1164 \\ 
\hline
1.4815 & 1.9431 & -0.3117 & 0.5231 & 1.8825 & -1.8317 \\ 
\hline
1.1663 & 1.9467 & -1.8564 & 1.6447 & 1.9112 & -0.6290 \\ 
\hline
0.8911 & 2.0018 & -0.7190 & 0.6395 & 1.9916 & -1.1617 \\ 
\hline
1.2837 & 1.9852 & -0.6655 & 0.6932 & 2.0128 & -1.5889 \\ 
\hline
0.7569 & 1.9771 & -0.9006 & 0.3673 & 2.1029 & -1.5779 \\ 
\hline
1.6313 & 2.0300 & -0.8589 & 0.9376 & 2.0127 & -2.0179 \\ 
\hline
0.7577 & 1.9821 & -0.5676 & 0.5676 & 1.9736 & -0.8992 \\ 
\hline
0.5482 & 2.0124 & -0.3588 & 0.7854 & 2.0058 & -0.0587 \\ 
\hline
1.2923 & 1.9270 & -0.3932 & 0.6114 & 1.8902 & -1.5024 \\ 
\hline
1.5750 & 1.9676 & -0.7003 & 0.8115 & 1.8921 & -1.7373 \\ 
\hline
0.9369 & 1.9325 & -0.5317 & 0.5394 & 1.9128 & -1.1926 \\ 
\hline
1.8380 & 2.0261 & -0.9785 & 0.9812 & 2.0308 & -2.3639 \\ 
\hline
1.2369 & 1.9782 & -0.6258 & 0.6211 & 2.0585 & -1.6346 \\ 
\hline
1.2424 & 2.0136 & -0.8368 & 0.8070 & 1.9979 & -1.5882 \\ 
\hline
0.6792 & 1.9826 & -0.4806 & 0.5497 & 1.9767 & -0.7058 \\ 
\hline
\end{tabular}
\caption{Boundaries gradients via Distorted GPR at 16 representative points.}
\label{table:DLSGPR}
\end{table}

Finally, we show in table \ref{table:ReprPoint} the percentages of observations represented by each of the 16 quantized points.

\begin{table}[h!]
\centering
\begin{tabular}{| c | c | c | c | c | c | c | c | c |} 
\hline $\mu_p$ & 
		0.0694 & 0.0208 & 0.0343 & 0.0167 & 0.0685 & 0.1428 & 0.0308 & 0.0119 \\
       $\%$ &
		0.70 & 0.76 & 3.53 & 10.49 & 1.79 & 0.72 & 5.66 & 14.99 \\
\hline $\mu_p$ & 
		0.0467 & 0.0165 & 0.0260 & 0.0130 & 0.0453 & 0.1002 & 0.0225 & 0.0088 \\
       $\%$ &
  	    2.11 & 11.22 & 5.52 & 12.11 & 2.87 & 1.19 & 8.33 & 11.00 \\
\hline
\end{tabular}
\caption{Percentage of observations represented by quantized point $\mu_p$.}
\label{table:ReprPoint}
\end{table}

\subsection{Implied Boundaries for Performance Measures}

\noindent Table \ref{table:QuantileRmupRep}-\ref{table:DLSGPRGammaRep} show boundaries for $\mu_p$, Sharpe ratio and acceptability index at the 16 quantized points. \footnote{The definition of Sharpe ratio adopted here is simply given by $\frac{\mu\sqrt{t}}{\sigma}$, with $\mu=\mu_p-\mu_n$, $\sigma^2=\sigma_p^2+\sigma_n^2$ and $t=250$ business days. The acceptability index is defined in \cite{MadanEberlein} as the maximal $\gamma$ such that the distorted expectation $\mathcal{E}^{\Psi_{\gamma}}[X]$ is nonnegative (or nonpositive for short position), where $\Psi_{\gamma}$ is again taken as the MINMAXVAR distortion.}

As observed in \cite{MadanEberlein}, typically acceptability indexes based on the MINMAXVAR distortion for returns on stocks and indexes are less than $0.15$, with median values of $0.04$ and more than 5\% of observations at $0$. These findings are consistent with the boundaries for the acceptability index at the 16 representative points shown in table \ref{table:QuantileGPRGammaRep} for both short and long positions. Note also that the acceptability index tends to be higher for long positions, which is also to be expected.

\begin{table}[h!]
\centering
 \begin{tabular}{| c | c | c || c | c | c |} 
 \hline
  \pbox{20cm}{ \ \ Upper \\ Boundary} 
 		& Observation
 		& \pbox{20cm}{ \ \ Lower \\ Boundary}
		& \pbox{20cm}{ \ \ Upper \\ Boundary} 
 		& Observation
 		& \pbox{20cm}{ \ \ Lower \\ Boundary}\\
	\hline\hline
 	0.0752 & 0.0694 & 0.0653 & 0.0512 & 0.0467 & 0.0427 \\ 
\hline
0.0228 & 0.0208 & 0.0189 & 0.0180 & 0.0165 & 0.0149 \\ 
\hline
0.0379 & 0.0343 & 0.0315 & 0.0287 & 0.0260 & 0.0238 \\ 
\hline
0.0177 & 0.0167 & 0.0157 & 0.0143 & 0.0130 & 0.0119 \\ 
\hline
0.0701 & 0.0685 & 0.0665 & 0.0470 & 0.0453 & 0.0432 \\ 
\hline
0.1459 & 0.1428 & 0.1402 & 0.1025 & 0.1002 & 0.0980 \\ 
\hline
0.0324 & 0.0308 & 0.0292 & 0.0238 & 0.0225 & 0.0212 \\ 
\hline
0.0127 & 0.0119 & 0.0109 & 0.0097 & 0.0088 & 0.0082 \\ 
\hline
\end{tabular}
\caption{$\mu_p$ boundaries estimated via Quantile Regression, at 16 representative points.}
\label{table:QuantileRmupRep}
\end{table}

\begin{table}[h!]
\centering
 \begin{tabular}{| c | c | c || c | c | c |} 
 \hline
  \pbox{20cm}{ \ \ Upper \\ Boundary} 
 		& Observation
 		& \pbox{20cm}{ \ \ Lower \\ Boundary}
		& \pbox{20cm}{ \ \ Upper \\ Boundary} 
 		& Observation
 		& \pbox{20cm}{ \ \ Lower \\ Boundary}\\
	\hline\hline
 	0.0856 & 0.0694 & 0.0697 & 0.0536 & 0.0467 & 0.0469\\
	\hline
    0.0224 & 0.0208 & 0.0190 & 0.0184 & 0.0165 & 0.0143\\
	\hline
    0.0348 & 0.0343 & 0.0339 & 0.0269 & 0.0260 & 0.0252\\
	\hline
    0.0180 & 0.0167 & 0.0153 & 0.0147 & 0.0130 & 0.0112\\
	\hline
    0.0706 & 0.0685 & 0.0661 & 0.0473 & 0.0453 & 0.0428\\
	\hline
    0.1440 & 0.1428 & 0.1421 & 0.1024 & 0.1002 & 0.0986\\
	\hline
    0.0329 & 0.0308 & 0.0284 & 0.0243 & 0.0225 & 0.0204\\
	\hline
    0.0127 & 0.0119 & 0.0107 & 0.0092 & 0.0088 & 0.0081\\
    \hline
\end{tabular}
\caption{$\mu_p$ boundaries estimated via Quantile GPR, at 16 representative points.}
\label{table:QuantileGPRmupRep}
\end{table}

\begin{table}[h!]
\centering
 \begin{tabular}{| c | c | c || c | c | c |} 
 \hline
  \pbox{20cm}{ \ \ Upper \\ Boundary} 
 		& Observation
 		& \pbox{20cm}{ \ \ Lower \\ Boundary}
		& \pbox{20cm}{ \ \ Upper \\ Boundary} 
 		& Observation
 		& \pbox{20cm}{ \ \ Lower \\ Boundary}\\
	\hline\hline
 	0.0756 & 0.0694 & 0.0644 & 0.0513 & 0.0467 & 0.0427 \\ 
\hline
0.0223 & 0.0208 & 0.0193 & 0.0175 & 0.0165 & 0.0154 \\ 
\hline
0.0377 & 0.0343 & 0.0317 & 0.0284 & 0.0260 & 0.0241 \\ 
\hline
0.0172 & 0.0167 & 0.0161 & 0.0137 & 0.0130 & 0.0123 \\ 
\hline
0.0699 & 0.0685 & 0.0676 & 0.0467 & 0.0453 & 0.0439 \\ 
\hline
0.1461 & 0.1428 & 0.1416 & 0.1025 & 0.1002 & 0.0992 \\ 
\hline
0.0320 & 0.0308 & 0.0297 & 0.0233 & 0.0225 & 0.0216 \\ 
\hline
0.0121 & 0.0119 & 0.0114 & 0.0090 & 0.0088 & 0.0086 \\ 
\hline
\end{tabular}
\caption{$\mu_p$ boundaries via Distorted LS at 16 representative points.}
\label{table:DLSLRmupRep}
\end{table}

\begin{table}[h!]
\centering
 \begin{tabular}{| c | c | c || c | c | c |} 
 \hline
  \pbox{20cm}{ \ \ Upper \\ Boundary} 
 		& Observation
 		& \pbox{20cm}{ \ \ Lower \\ Boundary}
		& \pbox{20cm}{ \ \ Upper \\ Boundary} 
 		& Observation
 		& \pbox{20cm}{ \ \ Lower \\ Boundary}\\
	\hline\hline
0.0751 & 0.0694 & 0.0694 & 0.0511 & 0.0467 & 0.0453 \\ 
\hline
0.0245 & 0.0208 & 0.0167 & 0.0182 & 0.0165 & 0.0143 \\ 
\hline
0.0409 & 0.0343 & 0.0274 & 0.0325 & 0.0260 & 0.0193 \\ 
\hline
0.0172 & 0.0167 & 0.0161 & 0.0138 & 0.0130 & 0.0120 \\ 
\hline
0.0694 & 0.0685 & 0.0664 & 0.0475 & 0.0453 & 0.0420 \\ 
\hline
0.1446 & 0.1428 & 0.1410 & 0.1021 & 0.1002 & 0.0982 \\ 
\hline
0.0324 & 0.0308 & 0.0284 & 0.0233 & 0.0225 & 0.0212 \\ 
\hline
0.0122 & 0.0119 & 0.0114 & 0.0091 & 0.0088 & 0.0086 \\ 
\hline

\end{tabular}
\caption{$\mu_p$ boundaries via Distorted GPR at 16 representative points.}
\label{table:DLSGPRmupRep}
\end{table}

%%%%%%%%%%%%%%%%%%%%%%%%%%%%%%% Sharpe %%%%%%%%%%%%%%%%%%%%%%%%%%%%%%%%%%%

\begin{table}[h!]
\centering
 \begin{tabular}{| c | c | c || c | c | c |} 
 \hline
  \pbox{20cm}{ \ \ Upper \\ Boundary} 
 		& Observation
 		& \pbox{20cm}{ \ \ Lower \\ Boundary}
		& \pbox{20cm}{ \ \ Upper \\ Boundary} 
 		& Observation
 		& \pbox{20cm}{ \ \ Lower \\ Boundary}\\
	\hline\hline
 	1.3983 & -0.1280 & -1.2170 & 1.0860 & -0.2864 & -1.5225 \\ 
\hline
1.6557 & 0.3176 & -0.9488 & 1.9242 & 0.6302 & -0.6800 \\ 
\hline
1.1930 & -0.2494 & -1.4179 & 1.4190 & -0.0292 & -1.1870 \\ 
\hline
2.9743 & 1.6717 & 0.3023 & 2.3051 & 0.9589 & -0.2970 \\ 
\hline
2.4155 & 0.9010 & -0.9568 & 2.0173 & 0.7184 & -0.9007 \\ 
\hline
2.5615 & 0.5177 & -1.2321 & 2.5085 & 0.6926 & -1.0699 \\ 
\hline
2.1170 & 0.7360 & -0.6640 & 2.4731 & 1.1418 & -0.2466 \\ 
\hline
3.1653 & 1.8893 & 0.5564 & 3.5950 & 2.2045 & 1.0172 \\ 
\hline
\end{tabular}
\caption{Sharpe ratio boundaries via Quantile Regression at 16 representative points.}
\label{table:QuantileRSharpeRep}
\end{table}

\begin{table}[h!]
\centering
 \begin{tabular}{| c | c | c || c | c | c |} 
 \hline
  \pbox{20cm}{ \ \ Upper \\ Boundary} 
 		& Observation
 		& \pbox{20cm}{ \ \ Lower \\ Boundary}
		& \pbox{20cm}{ \ \ Upper \\ Boundary} 
 		& Observation
 		& \pbox{20cm}{ \ \ Lower \\ Boundary}\\
	\hline\hline
 	4.0960 & -0.1253 & -0.0500 & 1.7966 & -0.2803 & -0.2231\\
 	\hline
    1.4038 &  0.3108 & -0.8727 & 2.2708 &  0.6168 & -1.1672\\
 	\hline
   -0.0615 & -0.2441 & -0.4023 & 0.4234 & -0.0286 & -0.4702\\
 	\hline
    3.2179 &  1.6361 & -0.1370 & 2.7574 &  0.9385 & -0.9590\\
 	\hline
    2.7685 &  0.8818 & -1.3389 & 2.1867 &  0.7031 & -1.2120\\
 	\hline
    1.2525 &  0.5067 &  0.0483 & 2.3867 &  0.6779 & -0.5442\\
 	\hline
    2.4903 &  0.7203 & -1.2574 & 2.9818 &  1.1174 & -0.9577\\
 	\hline
    3.0642 &  1.8490 &  0.2862 & 2.7892 &  2.1576 &  0.9789\\
 	\hline
\end{tabular}
\caption{Sharpe ratio boundaries via Quantile GPR, at 16 representative points.}
\label{table:QuantileGPRSharpeRep}
\end{table}

\begin{table}[h!]
\centering
 \begin{tabular}{| c | c | c || c | c | c |} 
 \hline
  \pbox{20cm}{ \ \ Upper \\ Boundary} 
 		& Observation
 		& \pbox{20cm}{ \ \ Lower \\ Boundary}
		& \pbox{20cm}{ \ \ Upper \\ Boundary} 
 		& Observation
 		& \pbox{20cm}{ \ \ Lower \\ Boundary}\\
	\hline\hline
1.5043 & -0.1280 & -1.4702 & 1.1167 & -0.2864 & -1.5371 \\ 
\hline
1.3725 & 0.3176 & -0.6988 & 1.4923 & 0.6302 & -0.3014 \\ 
\hline
1.1352 & -0.2494 & -1.3296 & 1.2542 & -0.0292 & -1.0275 \\ 
\hline
2.2277 & 1.6717 & 0.8526 & 1.6831 & 0.9589 & 0.2273 \\ 
\hline
2.1671 & 0.9010 & 0.0706 & 1.7605 & 0.7184 & -0.3756 \\ 
\hline
2.7175 & 0.5177 & -0.2952 & 2.4707 & 0.6926 & -0.1181 \\ 
\hline
1.7489 & 0.7360 & -0.2058 & 1.9404 & 1.1418 & 0.2373 \\ 
\hline
2.2607 & 1.8893 & 1.1881 & 2.4546 & 2.2045 & 1.8048 \\ 
\hline
\end{tabular}
\caption{Sharpe ratio boundaries via Distorted LS at 16 representative points.}
\label{table:DLSLRSharpeRep}
\end{table}

\begin{table}[h!]
\centering
 \begin{tabular}{| c | c | c || c | c | c |} 
 \hline
  \pbox{20cm}{ \ \ Upper \\ Boundary} 
 		& Observation
 		& \pbox{20cm}{ \ \ Lower \\ Boundary}
		& \pbox{20cm}{ \ \ Upper \\ Boundary} 
 		& Observation
 		& \pbox{20cm}{ \ \ Lower \\ Boundary}\\
	\hline\hline
1.3460 & -0.1253 & -0.1301 & 1.0333 & -0.2803 & -0.7017 \\ 
\hline
2.7489 & 0.3108 & -2.4183 & 2.0463 & 0.6168 & -1.2065 \\ 
\hline
2.3787 & -0.2441 & -3.0308 & 3.3902 & -0.0286 & -3.5272 \\ 
\hline
2.1957 & 1.6361 & 0.8460 & 1.7904 & 0.9385 & -0.1558 \\ 
\hline
1.6968 & 0.8818 & -1.0488 & 2.3775 & 0.7031 & -1.8278 \\ 
\hline
1.6345 & 0.5067 & -0.6340 & 2.1028 & 0.6779 & -0.9154 \\ 
\hline
2.0997 & 0.7203 & -1.2531 & 1.9408 & 1.1174 & -0.2121 \\ 
\hline
2.3064 & 1.8490 & 1.2318 & 2.5459 & 2.1576 & 1.7381 \\ 
\hline
\end{tabular}
\caption{Sharpe ratio boundaries via Distorted GPR at 16 representative points.}
\label{table:DLSGPRSharpeRep}
\end{table}

%%%%%%%%%%%%%%%%%%%%%%%%%%%%%%%%%%% Acceptability Index %%%%%%%%%%%%%%%%%%%%%%%%%%%

\begin{table}[h!]
\centering
 \begin{tabular}{| c | c | c || c | c | c |} 
 \hline
	\pbox{20cm}{ \ \ Upper \\ Boundary} 
 		& Observation
 		& \pbox{20cm}{ \ \ Lower \\ Boundary}
 		&\pbox{20cm}{ \ \ Upper \\ Boundary} 
 		& Observation
 		& \pbox{20cm}{ \ \ Lower \\ Boundary}\\
\hline\hline
0.0570 & 0.0036 & -0.0000 & 0.0457 & -0.0000 & 0.0000 \\ 
\hline
0.0667 & 0.0188 & 0.0000 & 0.0750 & 0.0283 & -0.0000 \\ 
\hline
0.0497 & -0.0000 & 0.0000 & 0.0578 & 0.0065 & 0.0000 \\ 
\hline
0.1132 & 0.0637 & 0.0123 & 0.0877 & 0.0394 & 0.0000 \\ 
\hline
0.0949 & 0.0378 & 0.0000 & 0.0809 & 0.0306 & 0.0000 \\ 
\hline
0.1027 & 0.0249 & -0.0000 & 0.1002 & 0.0309 & 0.0000 \\ 
\hline
0.0831 & 0.0303 & 0.0000 & 0.0957 & 0.0448 & -0.0000 \\ 
\hline
0.1207 & 0.0740 & 0.0210 & 0.1367 & 0.0858 & 0.0376 \\ 
\hline
\end{tabular}
\caption{Acceptability index boundaries via Quantile Regression at 16 representative points.}
\label{table:QuantileRGammaRep}
\end{table}

\begin{table}[h!]
\centering
 \begin{tabular}{| c | c | c || c | c | c |} 
 \hline
	\pbox{20cm}{ \ \ Upper \\ Boundary} 
 		& Observation
 		& \pbox{20cm}{ \ \ Lower \\ Boundary}
 		&\pbox{20cm}{ \ \ Upper \\ Boundary} 
 		& Observation
 		& \pbox{20cm}{ \ \ Lower \\ Boundary}\\
\hline\hline
0.1513 & -0.0039 & -0.0030 & 0.0646 & -0.0097 & -0.0087 \\ 
\hline
0.0506 & 0.0308 & -0.0109 & 0.0824 & 0.0413 & -0.0222 \\ 
\hline
-0.0150 & -0.0083 & -0.0017 & 0.0174 & -0.0153 & -0.0008 \\ 
\hline
0.1171 & 0.0588 & -0.0057 & 0.1004 & 0.0338 & -0.0338 \\ 
\hline
0.1010 & 0.0487 & -0.0322 & 0.0793 & 0.0440 & -0.0255 \\ 
\hline
0.0454 & 0.0188 & 0.0006 & 0.0870 & 0.0248 & -0.0203 \\ 
\hline
0.0904 & 0.0455 & -0.0261 & 0.1089 & 0.0403 & -0.0346 \\ 
\hline
0.1120 & 0.0674 & 0.0092 & 0.1017 & 0.0781 & 0.0341 \\ 
\hline
% 	 0.151 & -0.003 & -0.003 & 0.065 & -0.009 & -0.009 \\
% 	\hline
%     0.051 &  0.011 & -0.031 & 0.082 &  0.022 & -0.041 \\
% 	\hline
%    -0.015 & -0.008 & -0.002 & 0.015 & -0.001 & -0.017 \\
% 	\hline
%     0.117 &  0.059 & -0.006 & 0.100 &  0.034 & -0.034 \\
% 	\hline
%     0.101 &  0.032 & -0.049 & 0.079 &  0.026 & -0.044 \\
% 	\hline
%     0.045 &  0.019 &  0.001 & 0.087 &  0.025 & -0.020 \\
% 	\hline
%     0.090 &  0.026 &  0.046 & 0.109 &  0.040 & -0.035 \\
% 	\hline
%     0.112 &  0.067 &  0.009 & 0.102 &  0.078 &  0.034 \\
% 	\hline
\end{tabular}
\caption{Acceptability index boundaries via Quantile GPR at 16 representative points.}
\label{table:QuantileGPRGammaRep}
\end{table}

\begin{table}[h!]
\centering
 \begin{tabular}{| c | c | c || c | c | c |} 
 \hline
	\pbox{20cm}{ \ \ Upper \\ Boundary} 
 		& Observation
 		& \pbox{20cm}{ \ \ Lower \\ Boundary}
 		&\pbox{20cm}{ \ \ Upper \\ Boundary} 
 		& Observation
 		& \pbox{20cm}{ \ \ Lower \\ Boundary}\\
\hline\hline
0.0544 & 0.0036 & -0.0000 & 0.0408 & 0.0000 & 0.0000 \\ 
\hline
0.0504 & 0.0189 & -0.0000 & 0.0525 & 0.0288 & -0.0000 \\ 
\hline
0.0413 & 0.0000 & -0.0000 & 0.0456 & 0.0065 & 0.0000 \\ 
\hline
0.0801 & 0.0637 & 0.0365 & 0.0586 & 0.0395 & 0.0135 \\ 
\hline
0.0804 & 0.0378 & 0.0130 & 0.0650 & 0.0306 & -0.0000 \\ 
\hline
0.1017 & 0.0249 & 0.0013 & 0.0920 & 0.0308 & 0.0070 \\ 
\hline
0.0641 & 0.0305 & 0.0024 & 0.0702 & 0.0451 & 0.0166 \\ 
\hline
0.0813 & 0.0734 & 0.0493 & 0.0877 & 0.0856 & 0.0695 \\ 
\hline
\end{tabular}
\caption{Acceptability index boundaries via Distorted LS, at 16 representative points.}
\label{table:DLSLRGammaRep}
\end{table}

\begin{table}[h!]
\centering
 \begin{tabular}{| c | c | c || c | c | c |} 
 \hline
	\pbox{20cm}{ \ \ Upper \\ Boundary} 
 		& Observation
 		& \pbox{20cm}{ \ \ Lower \\ Boundary}
 		&\pbox{20cm}{ \ \ Upper \\ Boundary} 
 		& Observation
 		& \pbox{20cm}{ \ \ Lower \\ Boundary}\\
\hline\hline
0.0490 & -0.0058 & -0.0036 & 0.0369 & -0.0256 & -0.0099 \\ 
\hline
0.0995 & 0.0866 & -0.0107 & 0.0741 & 0.0427 & -0.0222 \\ 
\hline
0.1140 & -0.0838 & -0.0102 & 0.1328 & -0.1211 & -0.0027 \\ 
\hline
0.0796 & 0.0588 & 0.0290 & 0.0650 & 0.0337 & -0.0053 \\ 
\hline
0.0616 & 0.0382 & -0.0322 & 0.0861 & 0.0662 & -0.0255 \\ 
\hline
0.0593 & 0.0235 & -0.0188 & 0.0765 & 0.0335 & -0.0248 \\ 
\hline
0.0757 & 0.0454 & -0.0262 & 0.0701 & 0.0403 & -0.0083 \\ 
\hline
0.0835 & 0.0670 & 0.0423 & 0.0922 & 0.0777 & 0.0614 \\ 
\hline
\end{tabular}
\caption{Acceptability index boundaries via Distorted GPR, at 16 representative points.}
\label{table:DLSGPRGammaRep}
\end{table}

\subsection{Dimensional Analysis}
%
%In classical mean-variance analysis, the risks are all captured by a single parameter, and so the range of values of the reward for given volatility is relatively large, and risks/reward relationships
%are only assessed along the one dimensional efficient frontier. For a  three dimensional risk vector, it is, on the other hand, reasonable to ask whether observations lie on a (possibly noisy) lower dimensional embedding. In fact, the numerical results obtained from quantile regression show that the range of acceptable values for $\mu_p$ is relatively small compared to $\mu_p$ itself. 
%
As visible from tables \ref{table:QuantileRmupRep}-\ref{table:DLSGPRmupRep}, the different methodologies do not produce significantly different estimates for the boundaries $f_m$ and $f_M$, and one may wonder if such boundaries are indeed linear.  As the boundaries are close to each others one way to assess if this is the case is to compare the variance of the noise of a linear lower dimensional embedding with that of a nonlinear one.\footnote{For the nonlinear embedding we utilize the Diffusion map algorithm, recently introduced in \cite{CoifmanLafon}.} Our results, summarized in table, provide evidence of the linearity of $f_m$ and $f_M$. 
\begin{table}[h!]
\centering
 \begin{tabular}{|c || c | c | c | c|} 
 \hline & PCA
 		& cumulative weight (in $\%$) 
 		& Diffusion Map
 		& cumulative weight (in $\%$) \\
 \hline\hline
 $\lambda_1$ & 2.7529 & 68.82 & 0.0113 & 70.27\\ 
 \hline
 $\lambda_2$ & 1.1778 & 98.27 & 0.0045 & 98.58\\ 
 \hline
 $\lambda_3$ & 0.0685 & 99.98 & 0.0002 & 99.64\\ 
 \hline
 $\lambda_4$ & 0.0009 & 100.0 & 0.0001 & 100.0\\
 \hline
\end{tabular}
\caption{Eigenvalues's weights for PCA and diffusion map on the quantized dataset.}
\label{table:DimRedRes}
\end{table}

\section{A Simple Modification of a Lucas Tree Economy}

To formally link the risk-seeking behaviors observed above with those of prospects theory consider the following modification of a Lucas tree economy (\cite{Lucas1978}). There are two periods, and each agent is endowed with a single risky asset with payoff $S_i$ at the end of period $i$, $i=0,1$. Assume $S_1=S_0e^{G-L}$, where $G$ and $L$ are independent gamma distributed random variables. Suppose there is a risk-free asset in zero net supply with risk-free rate $r_{f}$, and that agents decide how mucht to borrow/lend at time $0$. Denoting such amount by $\ell$, consumption $C_i$ at period $i$, $i=0,1$, is
\begin{align*}
C_0=S_0+\ell, \
C_1=S_0e^{G-L}-\ell e^{r_f}.
\end{align*}
Finally, setting $X=G-L$ and $s_0=\log(S_0)$, suppose that, for some $0<\rho,\beta<1$, agents preferences are described by
\begin{align}\label{PTU}
U(C_0,C_1)=\frac{(\log(C_0))^{1-\rho}}{1-\rho}
+e^{-\beta}
\E\left[\frac{(\log(C_1))^{1-\rho}}{1-\rho}\indic{s_0+X\geq 0}
-\frac{(-\log(C_1))^{1-\rho}}{1-\rho}\indic{s_0+X\leq 0}\right].
\end{align}
This is a slight modification of the specification introduced in \cite{KahnemanTverski1992} to provide a working framework that includes prospects theory experimental observations. In particular, the investor is risk averse if and only if the log-return $G-L$ is above the threshold $s_0$, and this is a behavior that cannot be captured by preferences over terminal wealth.

\begin{figure*}
		\centering
        \begin{subfigure}[b]{0.46\textwidth}
            \centering
            \includegraphics[width=\textwidth]
            {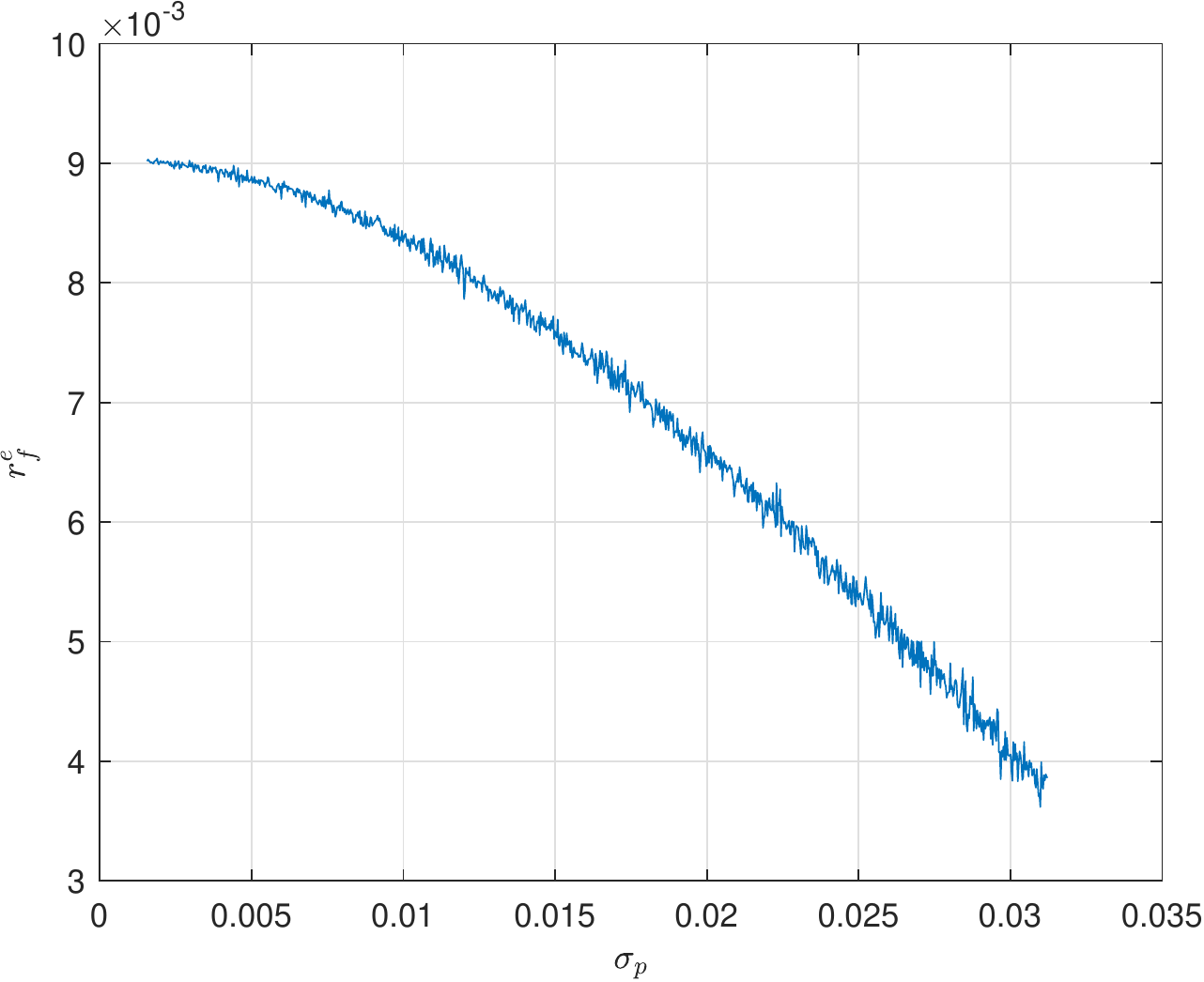}
            \caption[]%
            {{}}    
        \end{subfigure}
        \begin{subfigure}[b]{0.46\textwidth}
            \centering
            \includegraphics[width=\textwidth]
            {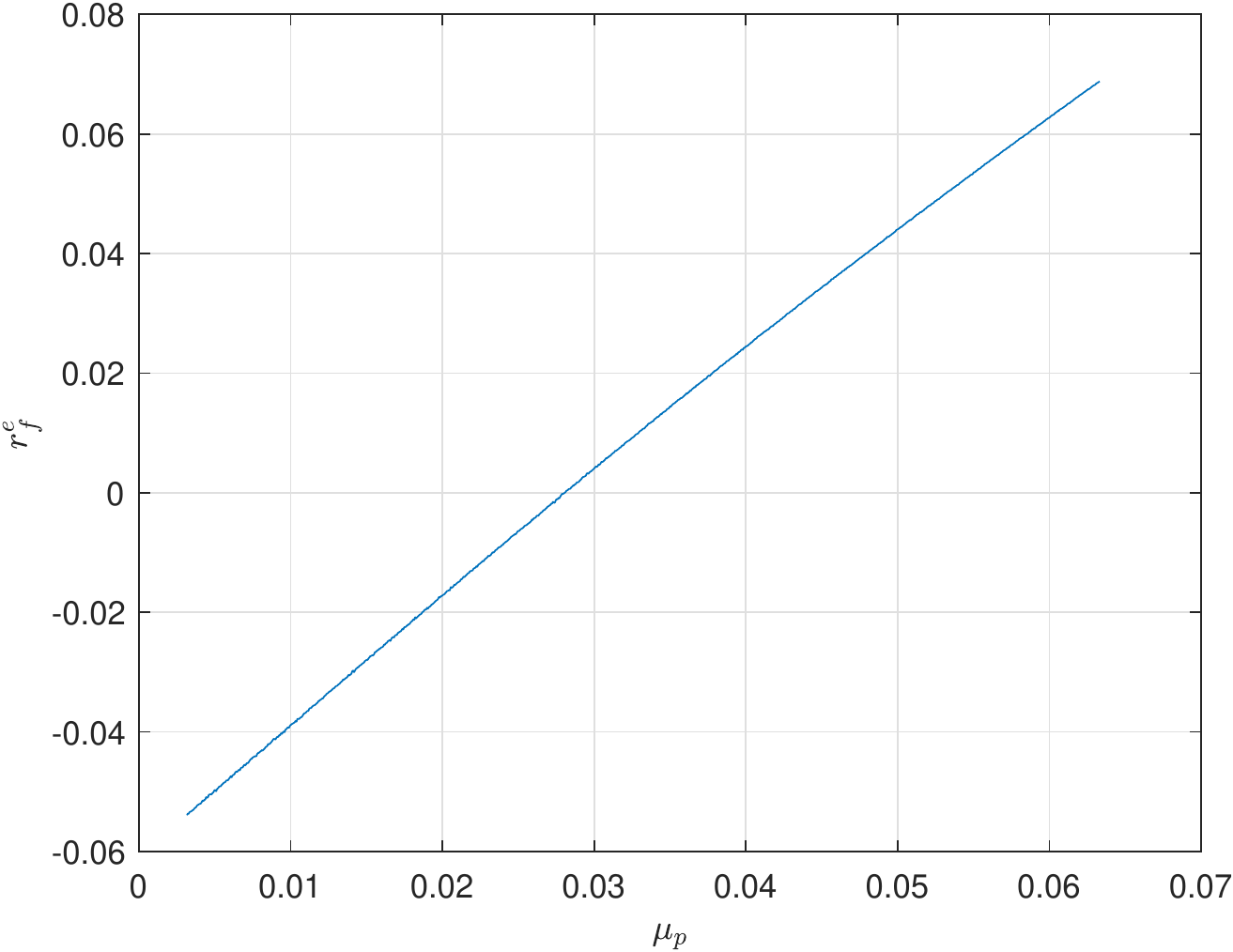}
            \caption[]%
            {{}}    
        \end{subfigure}
        \caption{Equilibrium rate as a function of $\sigma_p$ (a) and of $\mu_p$ (b).}\label{Eqp}
        \centering
        \begin{subfigure}[b]{0.46\textwidth}  
            \centering 
            \includegraphics[width=\textwidth]
            {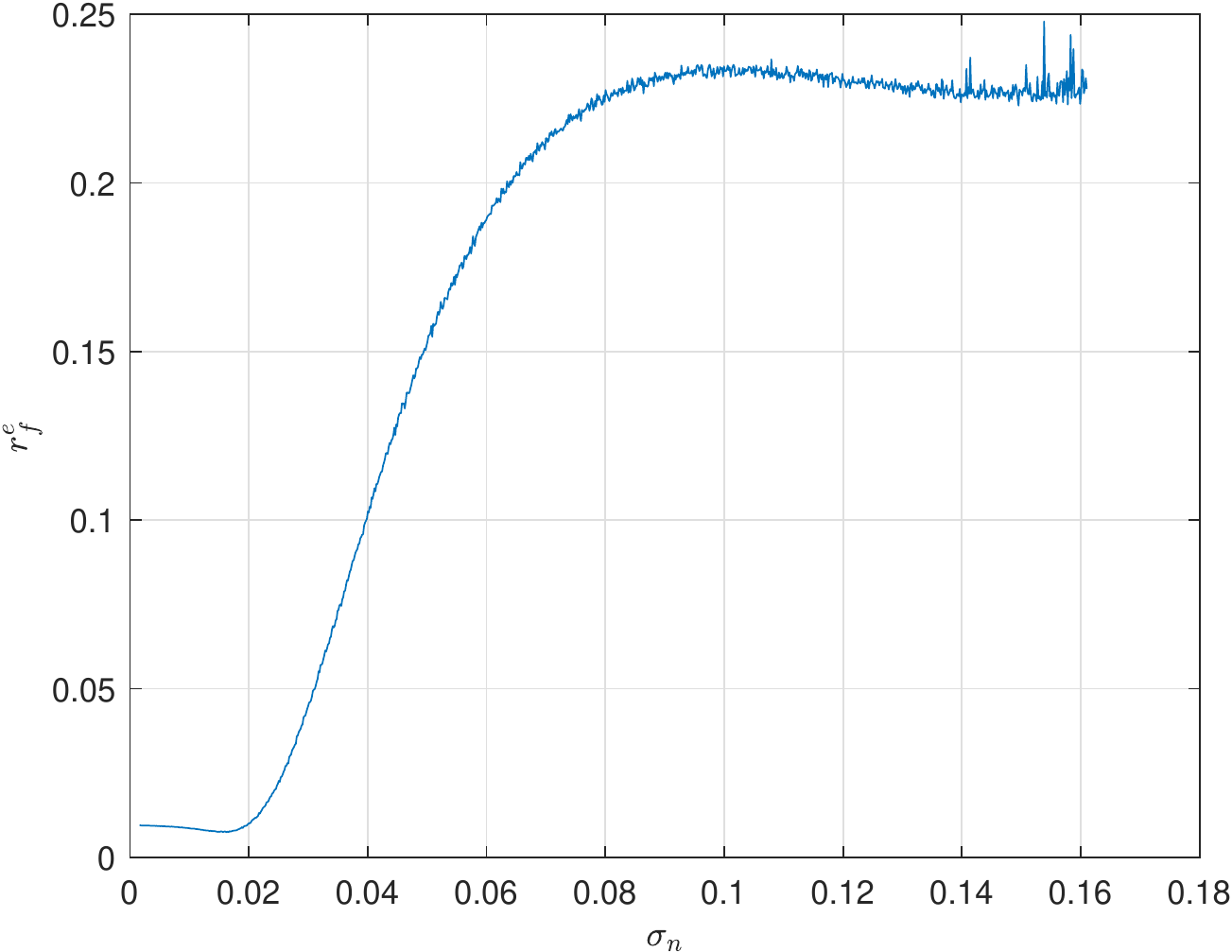}
            \caption[]%
            {{}}  
        \end{subfigure}
        \begin{subfigure}[b]{0.46\textwidth}  
            \centering 
            \includegraphics[width=\textwidth]
            {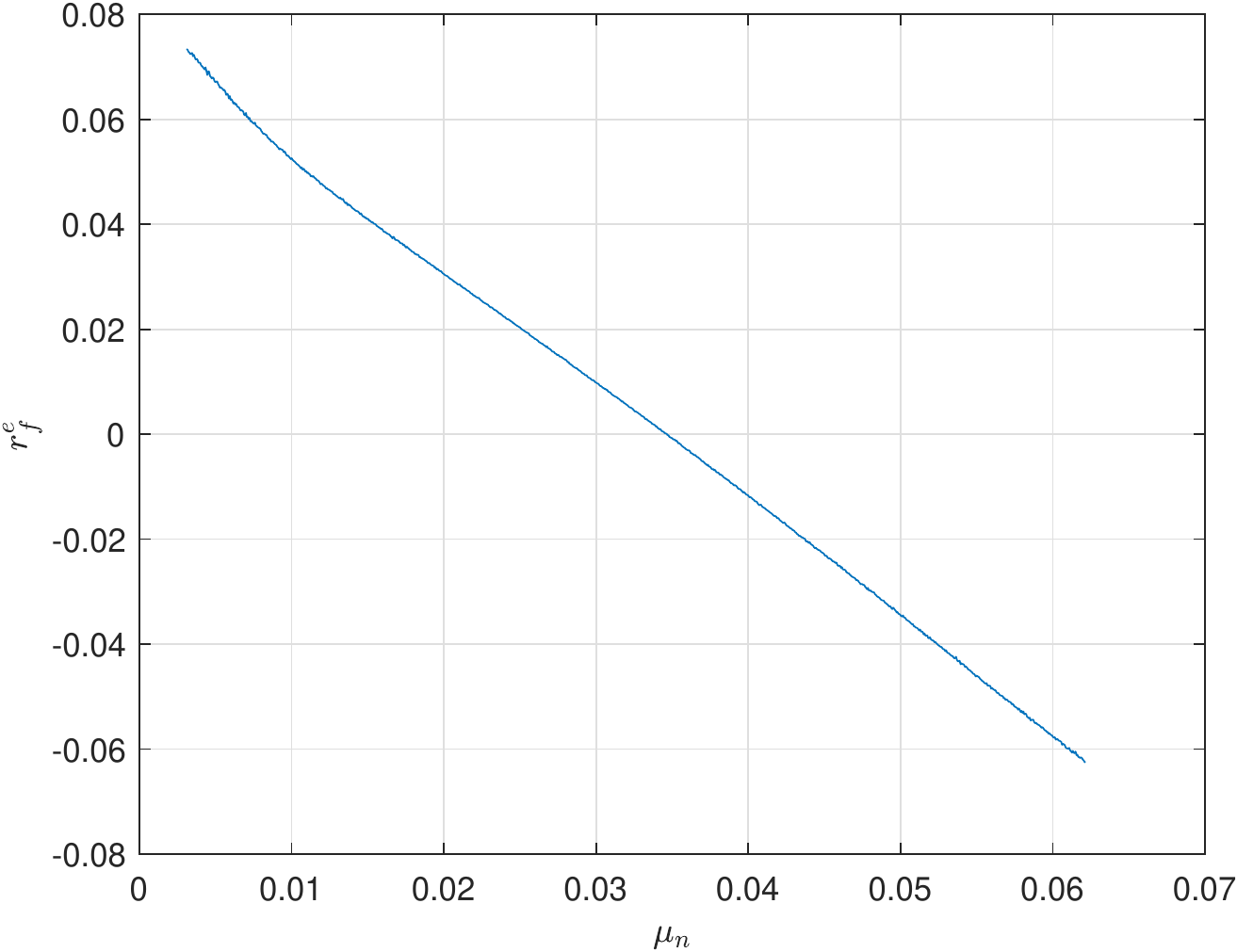}
            \caption[]%
            {{}}  
        \end{subfigure}
        \caption{Equilibrium rate as a function of $\sigma_n$ (a) and of $\mu_n$ (b).}\label{Eqn}
\end{figure*}

In equilibrium, $\ell=0$, and so 
\begin{align*}
s_0^{-\rho}
	-e^{r_f}e^{-\beta}\left(
	\E[(s_0+X)^{-\rho}e^{-X}\indic{s_0+X\geq 0}] 
	-\E[(-s_0-X))^{-\rho}e^{-X}\indic{s_0+X\leq 0}] 
	\right)
= 0,
\end{align*}
and so the equilibrium interest rate $r_f^e$ satisfies
\begin{align}\label{Equilibrium}
r_f^e=\beta-\rho\log(s_0)-\log\left(
	\E[(s_0+X)^{-\rho}e^{-X}\indic{s_0+X\geq 0}] 
	-\E[(-s_0-X))^{-\rho}e^{-X}\indic{s_0+X\leq 0}] 
	\right).
\end{align}
For a risk averse individual, higher risks correspond to lower equilibrium risk free rate, as lending becomes more attractive. Therefore, if the sign of $\partial r_f^e/\partial \sigma_n$ is negative, and that of $\partial r_f^e/\partial \sigma_p$ and $\partial r_f^e/\partial \mu_n$ are positive, the simple setting here described provides an explanation of our empirical findings. In general, it is possible to find values of $(\mu_p,\sigma_p,\mu_n,\sigma_n)$ and of $\rho$ such that this is indeed the case. For instance, setting $(\mu_p,\sigma_p,\mu_n,\sigma_n)=(0.03, 0.01,0.03, 0.01)$, which are the average values observed in the dataset above described, and setting $\rho=0.1$ and $\beta=0.01$, the value of $r_f^e$ computed via Montecarlo simulation as any of the variables $(\mu_p,\sigma_p,\mu_n,\sigma_n)$ changes is shown in figures \ref{Eqp} and \ref{Eqn}. As $\sigma_p$ and/or $\sigma_n$ increases the Montecarlo inegration estimate becomes less accurate as the variance of $X$ is higher, but the patterns in figures \ref{Eqp} and \ref{Eqn} confirm the behaviors above observed.

\section{The Risks-Neutral Acceptance Set}

In this section we analyze the ``risk-neutral'' acceptance set of quadruples $(b_p,c_p,b_n,c_n)$ of BG parameters estimated to option prices. The results reported are interesting on their own, but also test the methodology employed to analyze the acceptance set of statistical parameters.

\begin{figure*}
        \centering
        \begin{subfigure}[b]{0.45\textwidth}
            \centering
            \includegraphics[width=\textwidth]
            {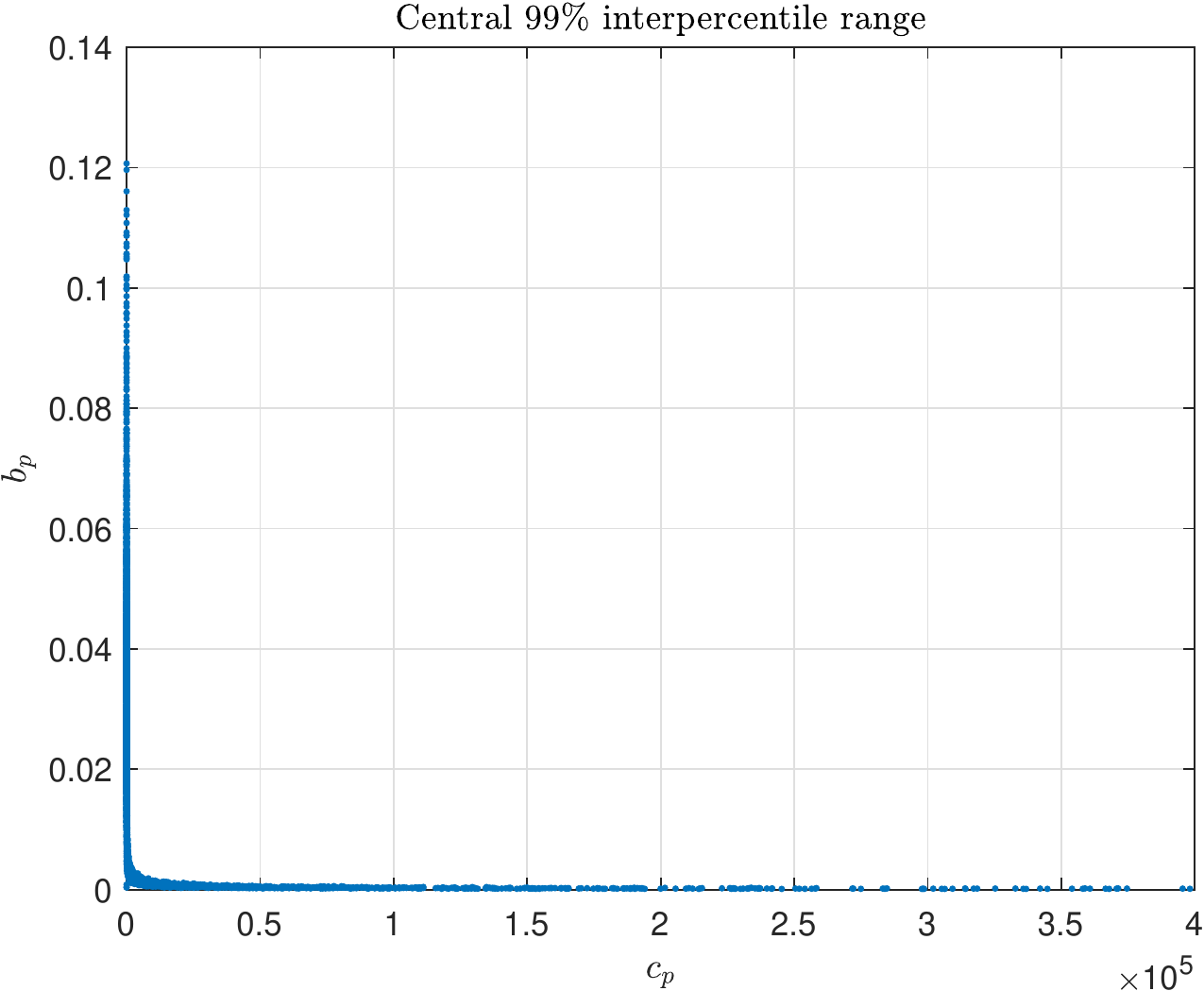}
            \caption[]%
            {{}}    
        \end{subfigure}
        \begin{subfigure}[b]{0.45\textwidth}  
            \centering 
            \includegraphics[width=\textwidth]
            {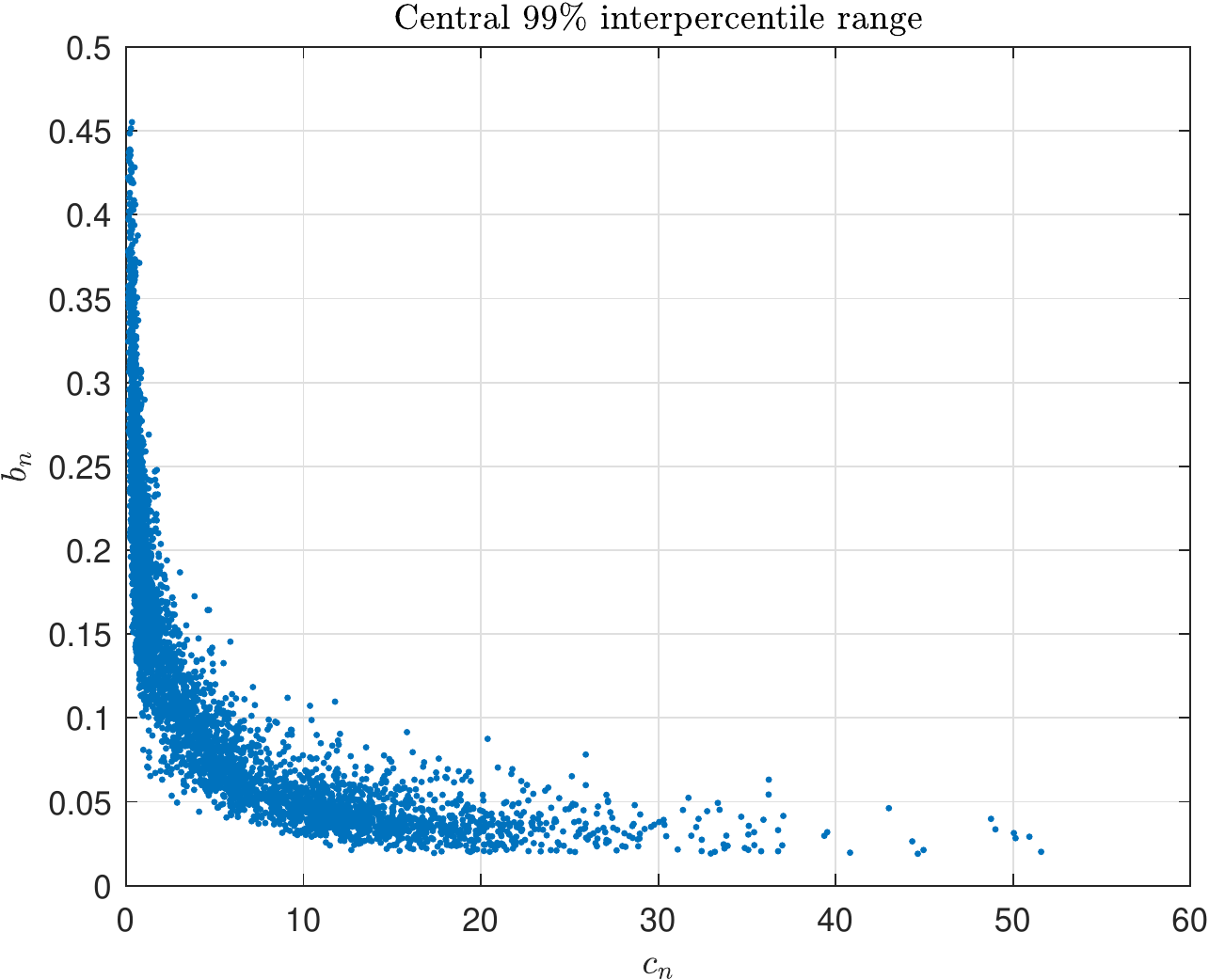}
            \caption[]%
            {{}}  
        \end{subfigure}
		\caption{Scatter plot of observed pairs $(b_p,c_p)$ and $(b_n,c_n)$ of risk neutral parameters.}\label{RNscatterdataset1}
        \begin{subfigure}[b]{0.45\textwidth}
            \centering
            \includegraphics[width=\textwidth]
            {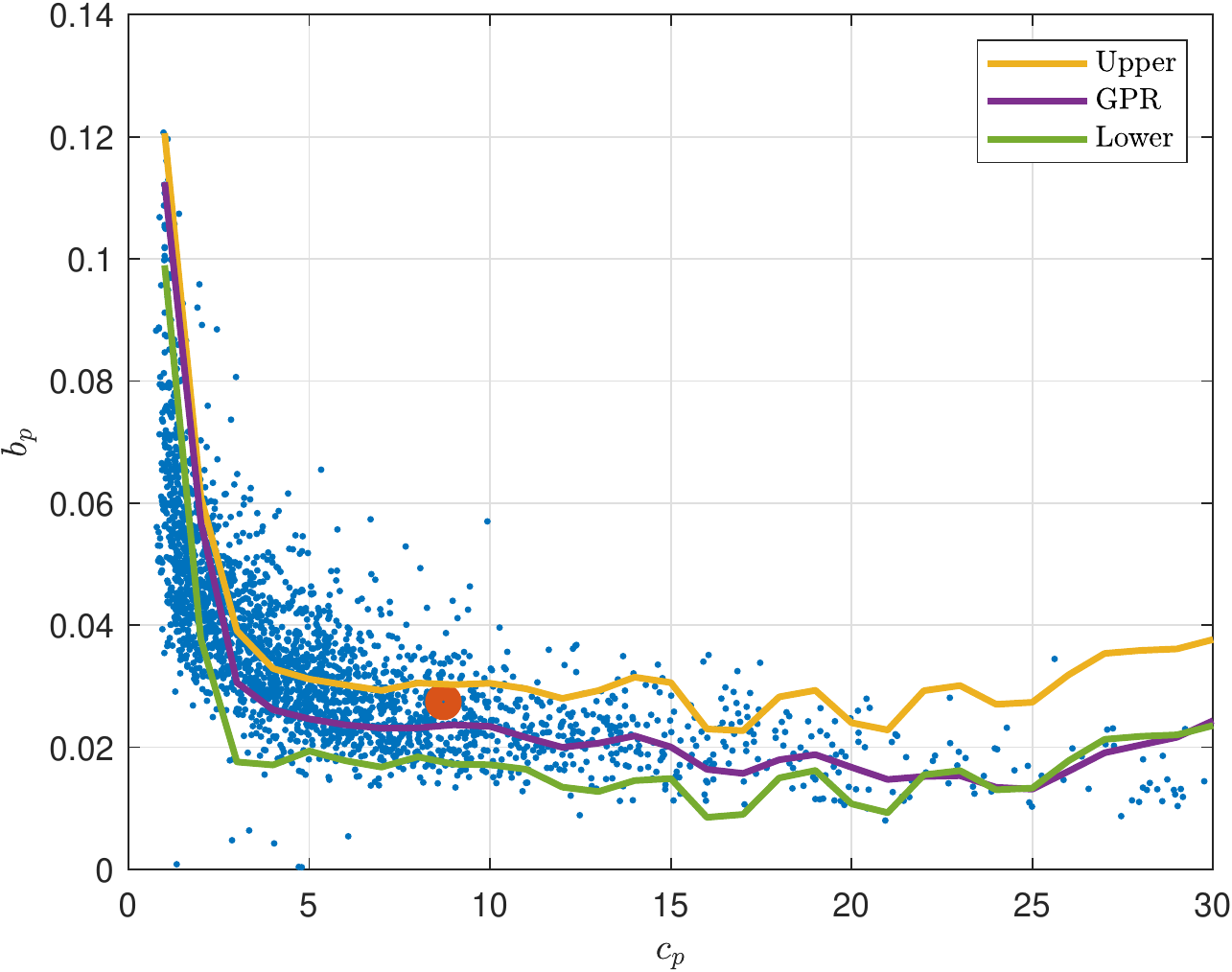}
            \caption[]%
            {{}}    
        \end{subfigure}
        \begin{subfigure}[b]{0.45\textwidth}  
            \centering 
            \includegraphics[width=\textwidth]
            {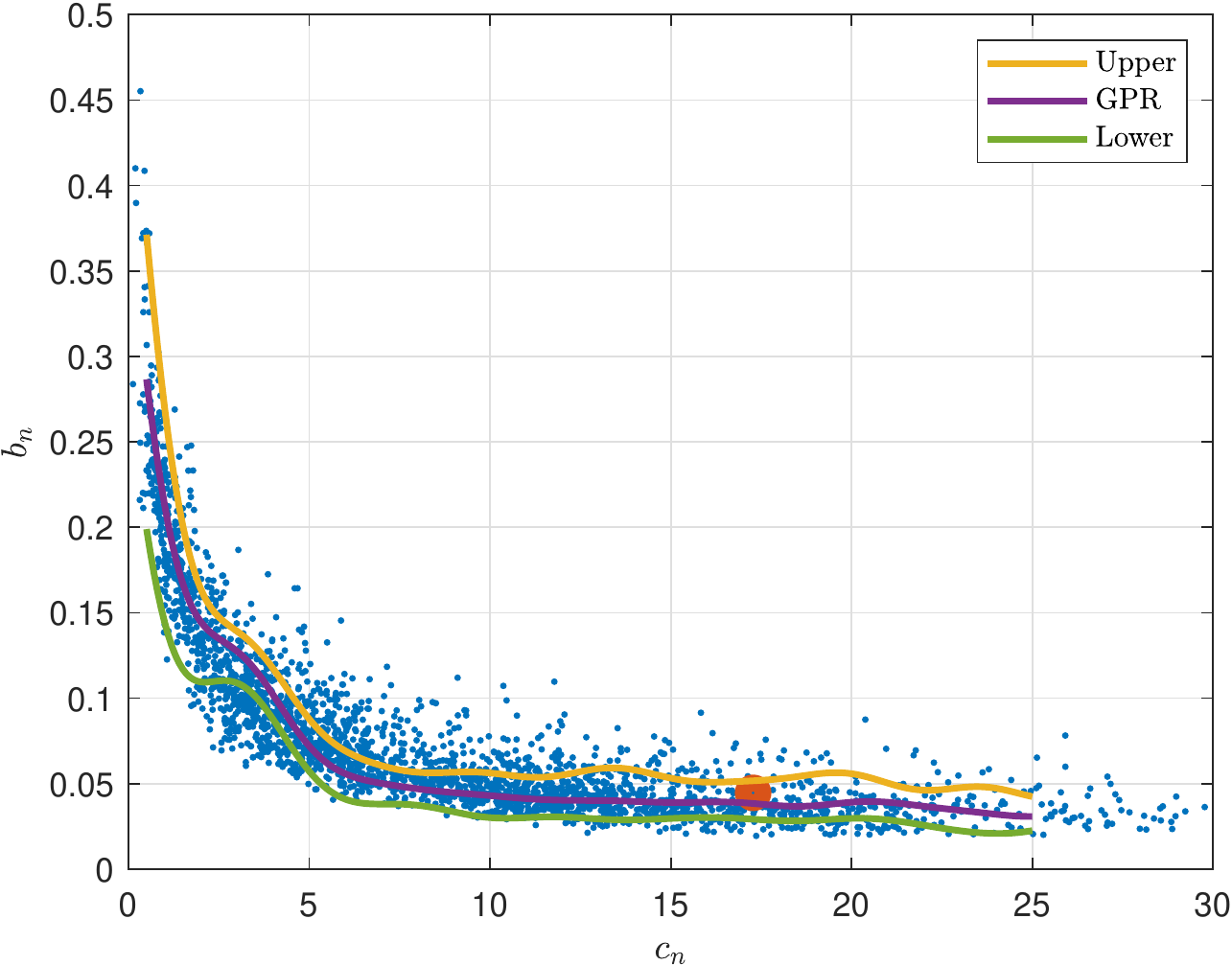}
            \caption[]%
            {{}}  
        \end{subfigure}
		\caption{Boundaries around randomly selected point (in red) via quantile GPR with ($\tau=0.05$).}\label{RNQGPR}
		\begin{subfigure}[b]{0.45\textwidth}
            \centering
            \includegraphics[width=\textwidth]
            {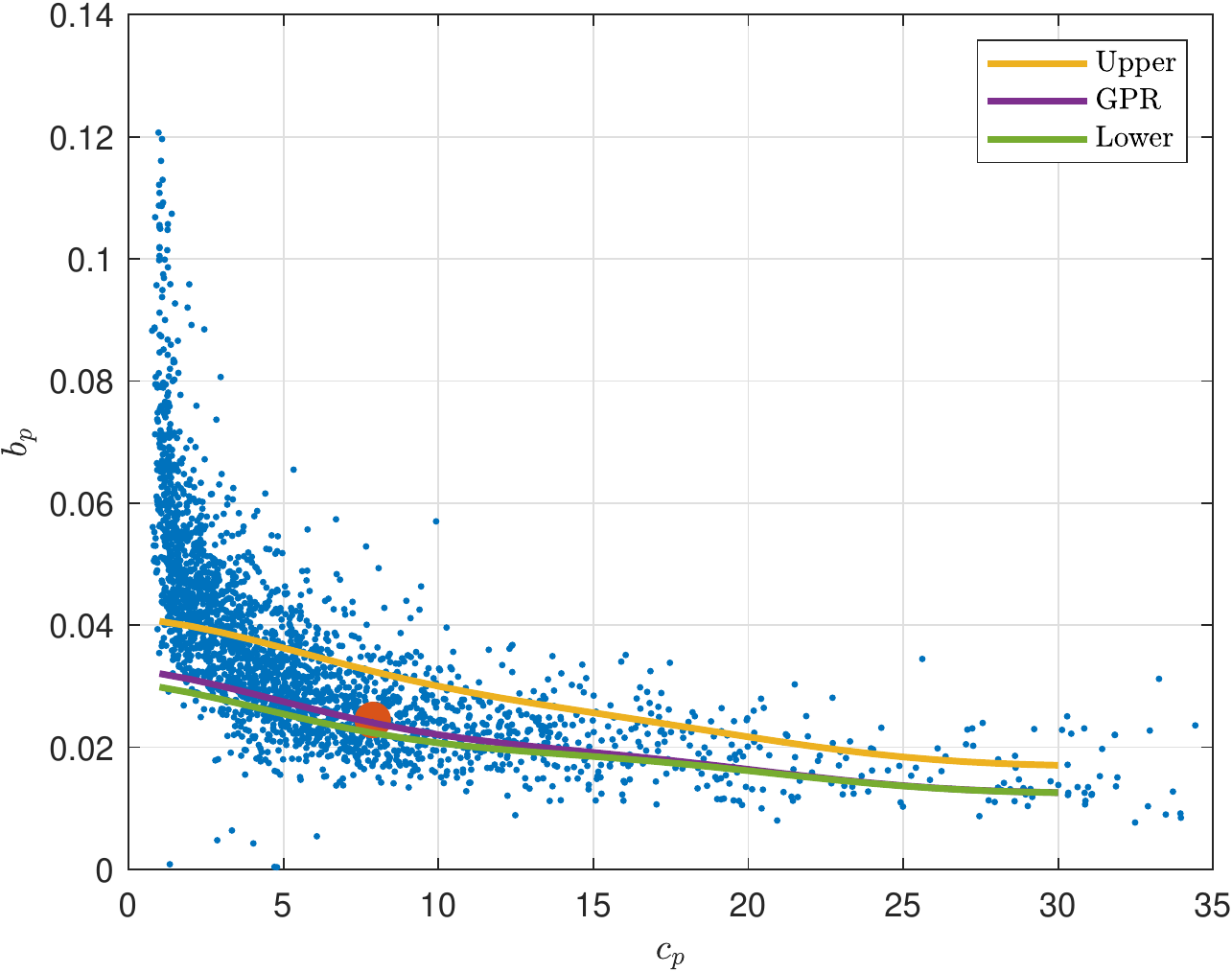}
            \caption[]%
            {{}}    
        \end{subfigure}
%        \begin{subfigure}[b]{0.34\textwidth}
%            \centering
%            \includegraphics[width=\textwidth]
%            {1scattersigma_p.eps}
%            \caption[]%
%            {{}}    
%        \end{subfigure}
        \begin{subfigure}[b]{0.45\textwidth}  
            \centering 
            \includegraphics[width=\textwidth]
            {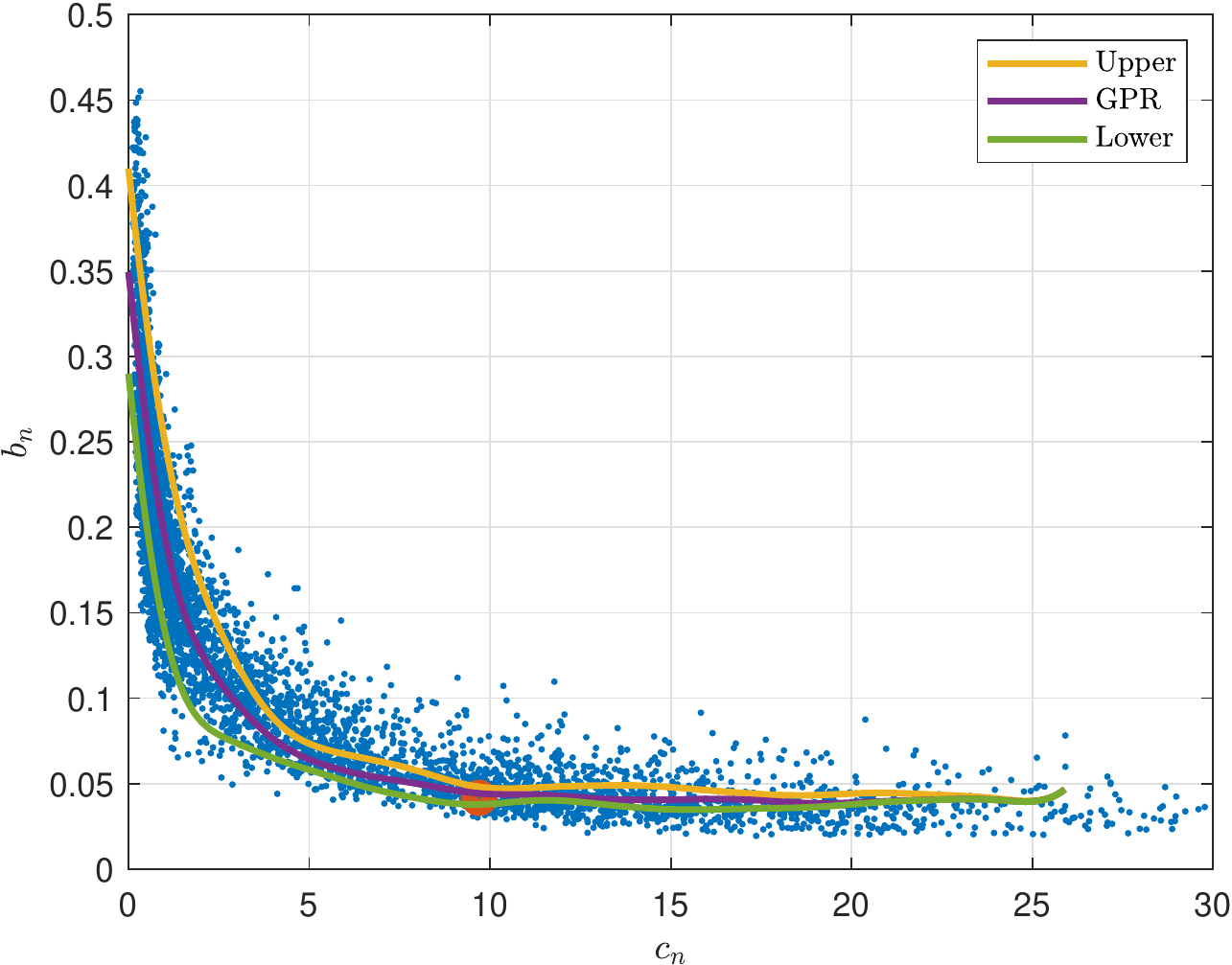}
            \caption[]%
            {{}}  
        \end{subfigure}
		\caption{Boundaries around randomly selected point (in red) via distorted GPR ($\gamma=0.75$).}\label{RNDLSGPR}
\end{figure*}

To better fit option prices, risk neutral log returns are modeled as $\omega t+X_t$, where $X_t$ is a BG process, $\omega := r+\log\left((1-b_p)^{c_p}(1+b_n)^{c_n}\right)$ and $r$ is the risk free rate. We calibrated the 10 sector ETFs to the mid prices of options for four different maturities\footnote{Of all the traded maturities, the middle four were considered. Tickers considered are SPY, XLB, XLE, XLF, XLI, XLK, XLP, XLU, XLV, XLY. Calibration was performed every 10 days between 1/01/2015 through 31/12/2020.}, and obtained a risk neutral dataset of 4812 observations. Figure \ref{RNscatterdataset1} shows pairs $(b_p,c_p)$ and $(b_n,c_n)$ excluding $1\%$ of outliers. Boundaries for $b_p$ and $b_n$ in terms of $(c_p,b_n,c_n)$ and $(c_p,b_n,c_n)$ are estimated via quantile\footnote{For quantile GPR, the optimization was performed employing a quasi-Newton method, with the quantile loss function approximated by $S(x)=\tau x+\alpha\log(1-e^{-x/\alpha})$ as in \cite{ZhengSmoothQR}, with $\alpha=10^{-4}$.} and distorted GPR. \footnote{In both cases, the hyperparameters of the kernel matrix $K$ are estimated using the standard MSE loss function, while $\alpha\in\R^n$ and $\beta\in\R$ are computed so that $\beta+K\alpha$ minimizes the quantile loss function.}
%\footnote{This methodology outperformed both transformation into a linear program and replacement of the constant basis in the GPR with the first 100 trigonometric basis or a feedforward neural network.}
Estimates are visualized in figure \ref{RNQGPR} and reported in table \ref{table:RNQGPRbp} and \ref{table:RNQGPRbn}.

We observe that both quantile and distorted regression tend to break down in estimating the boundaries of $b_p$ for large values of $c_p$, mostly because this parameter ranges between $0$ and $10^5$, with approximately $60\%$ of the observations concentrated in the range $[0,30]$ and the remaining ones being sparse (compare figure \ref{RNscatterdataset1}.A and \ref{RNQGPR}.A) and corresponding to relatively small variations in $(b_p,c_n,b_n)$. To avoid this issue, which - it is worth noting - does not occur when estimating boundaries of $b_n$ (note that the range of observations for $c_n$ is $[0,50]$), the regression algorithms for the boundaries of $b_p$ are only based on observations corresponding to $c_p<30$.

% For higher values of $c_p$, one may impose constraints on the optimization process based on the maximal and minimal values of $b_p$ observed for $c_p$ in small intervals, but this was not further investigated here. 

\begin{table}[h!]
\centering
 \begin{tabular}{| c | c | c || c | c | c |} 
 \hline
	\pbox{20cm}{ \ \ Upper \\ Boundary} 
 		& Observation
 		& \pbox{20cm}{ \ \ Lower \\ Boundary}
 		&\pbox{20cm}{ \ \ Upper \\ Boundary} 
 		& Observation
 		& \pbox{20cm}{ \ \ Lower \\ Boundary}\\
 \hline\hline
    0.0603 &  0.0451  & 0.0363  & 0.0304 &  0.0220  & 0.0136  \\
    \hline
    0.0684 &  0.0557  & 0.0370  & 0.0327 &  0.0265  & 0.0155  \\
    \hline
    0.0466 &  0.0382  & 0.0285  & 0.0307 &  0.0216  & 0.0165  \\
    \hline
    0.0385 &  0.0320  & 0.0248  & 0.0317 &  0.0215  & 0.0205  \\
    \hline
    0.0362 &  0.0296  & 0.0196  & 0.0393 &  0.0313  & 0.0244  \\
    \hline
    0.0317 &  0.0243  & 0.0189  & 0.0282 &  0.0202  & 0.0145  \\
    \hline
    0.0356 &  0.0276  & 0.0229  & 0.0263 &  0.0192  & 0.0125  \\
    \hline
    0.0329 &  0.0248  & 0.0180  & 0.0291 &  0.0197  & 0.0152  \\
    \hline
\end{tabular}
\caption{Boundaries for $b_p$ via quantile GPR at 16 representative points (with $c_p<30$).}
\label{table:RNQGPRbp}
\end{table}

\begin{table}[h!]
\centering
 \begin{tabular}{| c | c | c || c | c | c |} 
 \hline
	\pbox{20cm}{ \ \ Upper \\ Boundary} 
 		& Observation
 		& \pbox{20cm}{ \ \ Lower \\ Boundary}
 		&\pbox{20cm}{ \ \ Upper \\ Boundary} 
 		& Observation
 		& \pbox{20cm}{ \ \ Lower \\ Boundary}\\
 \hline\hline
	0.0728 & 0.0593 & 0.0430 & 0.2394 & 0.1862 & 0.1266 \\ 
\hline
0.0541 & 0.0346 & 0.0261 & 0.2422 & 0.1885 & 0.1284 \\ 
\hline
0.2576 & 0.1992 & 0.1347 & 0.2378 & 0.1852 & 0.1268 \\ 
\hline
0.2392 & 0.1857 & 0.1261 & 0.2198 & 0.1701 & 0.1193 \\ 
\hline
0.2597 & 0.2012 & 0.1356 & 0.2175 & 0.1674 & 0.1194 \\ 
\hline
0.2452 & 0.1906 & 0.1291 & 0.2259 & 0.1756 & 0.1215 \\ 
\hline
0.2703 & 0.2089 & 0.1406 & 0.2113 & 0.1679 & 0.1214 \\ 
\hline
0.2638 & 0.2041 & 0.1374 & 0.2302 & 0.1764 & 0.1280 \\ 
\hline
\end{tabular}
\caption{Boundaries for $b_n$ via quantile GPR at 16 representative points.}
\label{table:RNQGPRbn}
\end{table}

\subsection{Speed Uncertainty}
As mentioned, scale and shape parameters represent, respectively, limit and market orders. Typically, professional traders place limit orders based on stable patterns and strategies, and so the scale parameters arguably represent the phase of the economic cycle, and so the speed parameters can be thought of as noisy responses to it.\footnote{Diffusion map showed that more than $95\%$ of the dataset variance is explained by two eigenvectors.} This would be, however, outside of the theory of risk measures, as changing measure does not change speed parameters. Thus, theoretical boundaries similar to those derived in the next section would require a notion of ``speed uncertainty'', similar to that of volatility uncertainty of $G$-Brownian motion (\cite{Peng2006}). Such notion can be implemented via nonlinear Levy processes (\cite{NeufeldNutz}), according to which, e.g., the ask price of a claim $C=f(X_T)$, where $X$ is a bilateral gamma process, is the unique viscosity solutions of
\begin{align*}
\begin{cases}
ru(t,x)+\sup_{c_p,c_n\in \Theta} \left \lbrace
	\int_{\R\setminus\{0\}}[u(t,x+y)-u(t,x)]k(y)dy\right\rbrace=u_t(t,x),\\
u(0,x)=f(x)
\end{cases}
\end{align*}
where $\Theta\subset \R^2$ is compact. There is however a large literature on the magnitude of the spread between upper and lower valuations based on spectral risk measure, and since our empirical analysis in the next sections is based on it, this approach was not further investigated.

\subsection{An Equation for the Boundaries of Acceptable Risk Neutral Parameters}\label{BoundaryEquation}

As mentioned in the introduction, the boundaries found for the risk neutral parameters are naturally linked to acceptance sets implied by risk measures. In particular, given a fixed probability space $(\Omega,\mathcal{F},\P)$ and an asset's bid and ask price processes $\{B_t\}_{t\geq 0}$ and $\{A_t\}_{t\geq 0}$, a version of the first fundamental theorem of asset pricing with transaction costs asserts the existence of a probability measure $\Q$ and a processes $\{S_t\}_{t\geq 0}$ such that $\Q$ is equivalent to $\mathcal{\P}$, $B_t\leq S_t\leq A_t$ for every $t\geq 0$ and $\{e^{-rt}S_t\}_{t\geq 0}$ is a martingale under $\Q$.\footnote{For the existence of $\Q$ and the associated processes $\{S_t\}_{t\geq 0}$ see \cite{Jouini} and \cite{SchachermeyerFTC}.} Note that the measure $\Q$ is, approximately, a risk neutral measure in the sense that the process $S_t$ approximates the price at which one can buy and sell the asset. One can then assume that the asset's bid and ask prices be given by
\begin{equation*}
\begin{aligned}
B_0 = \inf_{Q\in \mathcal{M}}\E^{Q}[S_0e^{-r+\omega+X_1}], \
A_0 = \sup_{Q\in \mathcal{M}}\E^{Q}[S_0e^{-r+\omega+X_1}].
\end{aligned}
\end{equation*}
where $\mathcal{M}$ is a collection of probability measures that are equivalent to the statistical measure $\P$. Such collection is a financial primitive of the economy that, as anticipated in the introduction, depends on regulator's requirements for financial stability as well as trading, costs and incentives of market operators, and a risk $Z$ is deemed acceptable if $\E^{Q}[Z]\geq 0, \forall Q\in\mathcal{M}$.
 %As in the previous section it is assumed that, for $t\geq 0$, $S_t=e^{\omega t+X_t}$ with $\{X_t\}_{t\geq 0}$ a BG process under $\Q$ and $\omega$ is defined by \ref{omega}. 
For our purposes, $\mathcal{M}$ is defined as the set of measures associated to the spectral risk measure that arise from a distortion $\Psi$ ((one can employ e.g. the MINMAXVAR defined by \ref{MINMAXVAR}). Bid and ask prices are then computed as integrals of distorted probabilities of tail events (see \cite{MDV}), and the higher their distortion the higher size of the set $\mathcal{M}$ and the bid-ask spread.

Next, suppose that for given risk neutral parameters $(\hat{c}_p,\hat{b}_n,\hat{c}_n)$, the corresponding $b_p$ lies in the interval $[\underline{b}_p,\overline{b}_p]$. It is natural to assume that
\begin{align}\label{AcceptabilityCondition}
\{\Q_{b_p}\}_{b_p\in[\underline{b}_p,\overline{b}_p]}\subset \mathcal{M}.
\end{align}
where, for every $b_p\in [\underline{b}_p,\overline{b}_p]$, $\Q_{b_p}$ is a measure under which $\{X_t\}_{t\geq 0}$ is a BG process with parameters $(b_p,\hat{c}_p,\hat{b}_n,\hat{c}_n)$. Note that, by proposition 6.1 in \cite{KuchlerTappe}, such a measure exists and is equivalent to the risk neutral measure $\Q$ (and thus also to the statistical measure $\P$). Since
\begin{equation}\label{alphabeta1dollar}
\begin{aligned}
\inf_{b_p\in [\underline{b}_p,\overline{b}_p]}
	\E^{\Q_{b_p}}[e^{-r+\omega+X_1}]
	= \frac{(1-\hat{b}_p)^{\hat{c}_p}}{(1-\underline{b}_p)^{\hat{c}_p}},\
\sup_{b_p\in [\underline{b}_p,\overline{b}_p]}
	\E^{\Q_{b_p}}[e^{-r+\omega+X_1}]
	= \frac{(1-\hat{b}_p)^{\hat{c}_p}}{(1-\overline{b}_p)^{\hat{c}_p}},
\end{aligned}
\end{equation}
where $\omega=r+\log((1-\hat{b}_p)^{\hat{c}_p}(1+\hat{b}_n)^{\hat{c}_n})$, \ref{AcceptabilityCondition} implies
\begin{equation}
\begin{aligned}
\frac{B_0}{S_0}
\leq  \frac{(1-\hat{b}_p)^{\hat{c}_p}}{(1-\underline{b}_p)^{\hat{c}_p}}
\leq \frac{(1-\hat{b}_p)^{\hat{c}_p}}{(1-\overline{b}_p)^{\hat{c}_p}}
\leq \frac{A_0}{S_0}.
\end{aligned}
\end{equation}
Further pushing \ref{AcceptabilityCondition} to be satisfied with an equality, we obtain the following relation for the upper and lower boundaries $\overline{b}_p$ and $\underline{b}_p$:
\begin{equation}\label{RNBoundary}
\frac{B_0}{A_0}
=  \frac{(1-\overline{b}_p)^{\hat{c}_p}}{(1-\underline{b}_p)^{\hat{c}_p}}
\end{equation}
Note that, in \ref{RNBoundary}, $B_0$ and $A_0$ are functions of $\hat{c}_p,\hat{b}_n,\hat{c}_n$. In other words, the parameters $\hat{c}_p,\hat{b}_n,\hat{c}_n$ are measures of economic activity and thus, together with the structural limits $\underline{b}_p,\overline{b}_n$, determine bid and ask prices. Similarly, if $\hat{c}_p$ and $\hat{c}_n$ determine boundaries for $b_p$ and $b_n$,
\begin{equation}\label{RNBoundary2}
\frac{B_0(\hat{c}_p,\hat{c}_n)}{A_0(\hat{c}_p,\hat{c}_n)}
=  \frac{(1-\overline{b}_p)^{\hat{c}_p}}{(1-\underline{b}_p)^{\hat{c}_p}}
	\frac{(1+\underline{b}_n)^{\hat{c}_n}}{(1+\overline{b}_n)^{\hat{c}_n}}.
\end{equation}

%If the bid-ask ratio remains constant with respect to any of the values of $(c_p,b_n,c_n)$, \ref{RNBoundary} explains the inverse relationship between upper jumps scale and shape parameters suggested by figure \ref{RNscatterdataset1}, as an increase in, say, $\hat{c}_p$ must be offset by a decrease in $\overline{b}_p$ or in $\hat{b}_p$. 

\subsection{Empirical Verifications}
Typically, equations \ref{RNBoundary} and/or \ref{RNBoundary2} are not satisfied, at least with respect to daily closing bid-ask ratios. However, since large orders are executed over several days, one can consider other distorted valuations, such as 5-days high/low prices. In general, one can compare the size of $\mathcal{M}$ required for \ref{AcceptabilityCondition} to hold with typically observed acceptability indexes. To do so, we let $\nu_{b_p}$ denote the BG Levy measure with parameters $(b_p,\hat{c}_p,\hat{b}_n,\hat{c}_n)$, and replace \ref{AcceptabilityCondition} with
\begin{align}\label{AcceptabilityConditionFrequency}
\{\nu_{b_p}\}_{b_p\in[\underline{b}_p,\overline{b}_p]}\subset \mathcal{N},
\end{align}
The collection $\mathcal{N}$ is such that distorted rewards are defined by
\begin{align*}
\underline{\mu} & = \omega-\int_0^{\infty}G^+(\nu(e^x-1<-a))da+\int_{0}^{\infty}(G^-(\nu(e^x-1>a))da, \\
\overline{\mu} &= \omega-\int_0^{\infty}G^-(\nu(e^x-1<-a))da+\int_{0}^{\infty}(G^+(\nu(e^x-1>a))da
\end{align*}
where $\nu$ is the Levy measure of $X$ under $\Q$ and $G^+$ and $G^-$ are (see \cite{EMPY})
\begin{align*}
G^+(x)=x+\frac{1}{c}(1-e^{-cx})^{1/(1+\gamma)}, \
G^-(x)=x-\frac{1}{c}(1-e^{-cx}).
\end{align*}
%There are at least two advantages of focusing on distorting arrival rates. Firstly, in the BG setting, Levy measures are easily obtained, while probability densities and distributions are more difficult to compute, and thus prone to generating higher numerical errors. Secondly, focusing on arrival rates allows one to deal with the issues of fixing the investment horizon, such as the fact that probabilities of tail events converge to zero as this becomes shorter (while arrival rates do not).
Then, as proved in \cite{MDV}, $\tilde{\nu}\in \mathcal{N}$ if and only if $\frac{d\tilde{\nu}}{d\nu}$ satisfies
\begin{equation}\label{AcceptabilityConditionMeasure}
\begin{aligned}
S(\lambda):&=\int_{\R}\left(\frac{d\tilde{\nu}}{d\nu}-\lambda\right)^+d\nu(x)\leq 
	\Phi(\lambda), \ \lambda>1,\\
\tilde{S}(\lambda):&=\int_{\R}\left(\lambda-\frac{d\tilde{\nu}}{d\nu}\right)^+d\nu(x)\leq 
	-\tilde{\Phi}(\lambda), \ 0\leq \lambda< 1,
\end{aligned}
\end{equation}
where $\Phi$ and $\tilde{\Phi}$ are Fenchel conjugates of $G^+$ and $G^-$ respectively, and are given by
\begin{align*}
\begin{aligned}
\Phi(\lambda) 
&:= \frac{1}{c}\left[-(1-\lambda)\log(u(\lambda))
	+\left(1-u(\lambda)\right)^{1/(1+\gamma)}\right],\\
-\tilde{\Phi}(\lambda) 
&:= \frac{1}{c}[\lambda+(1-\lambda)\log(1-\lambda)],
\end{aligned}
\end{align*}
with $u:(1,\infty)\rightarrow (0,1)$ defined as the unique solution of
\begin{align*}
\frac{u}{(1-u)^{\gamma/(1+\gamma)}}=(\lambda-1)(1+\gamma).
\end{align*}
Next, we find requirements on $c,\gamma$ for \ref{AcceptabilityConditionFrequency} to hold. Note that, for $\tilde{\nu}=\nu_{\overline{b}_p}$,
\begin{align*}
S(\lambda)
&=\int_{L(\lambda)}^{\infty} \frac{c_p}{x}\left(e^{-x/\overline{b}_p}-\lambda e^{-x/\hat{b}_p}\right)dx\
 =c_p \left(Ei\left(\frac{L(\lambda)}{\overline{b}_p}\right)
	- \lambda Ei\left(\frac{L(\lambda)}{\hat{b}_p}\right)\right)
\end{align*}
and, similarly, for $\tilde{\nu}=\nu_{\underline{b}_p}$,
\begin{align*}
\tilde{S}(\lambda)
&=\int_{\tilde{L}(\lambda)}^{\infty} \frac{c_p}{x}\left(\lambda e^{-x/\hat{b}_p}-e^{-x/\underline{b}_p}\right)dx\
 =c_p \left(\lambda Ei\left(\frac{L(\lambda)}{\hat{b}_p}\right)
	- Ei\left(\frac{L(\lambda)}{\underline{b}_p}\right)\right),
\end{align*}
where $Ei$ is the exponential integral function and 
\begin{align*}
L(\lambda)
 =\frac{\log(\lambda)\overline{b}_p\hat{b}_p}{\overline{b}_p-\hat{b}_p}, \
\tilde{L}(\lambda)
 =-\frac{\log(\lambda)\hat{b}_p\underline{b}_p}{\hat{b}_p-\underline{b}_p}.
\end{align*}
\begin{lemma} Suppose $\underline{b}_p>0.55\hat{b}_p$. Then, $\nu_{b_p}\in\mathcal{N}$ holds for every $b_p\in [\underline{b}_p,\hat{b}_p]$ if and only if
\begin{align}\label{ConditionLower}
c \leq \lim_{\lambda\rightarrow 1^-}\frac{1}{\tilde{S}(\lambda)}
\end{align}
\end{lemma}
\begin{proof} Note that for every $0<\lambda<1$
\begin{align*}
-\tilde{\Phi}'(\lambda)-\tilde{S}'(\lambda)
&=\frac{-\log(1-\lambda)}{c}
 -c_p \left(Ei\left(\frac{L(\lambda)}{\hat{b}_p}\right)
 	-\lambda e^{-L(\lambda)/\hat{b}_p}\frac{L'(\lambda)}{L(\lambda)}
	+ e^{-L(\lambda)/\underline{b}_p}\frac{L'(\lambda)}{L(\lambda)}\right)\\
&=\frac{-\log(1-\lambda)}{c}
 -c_p Ei\left(\frac{L(\lambda)}{\hat{b}_p}\right),
\end{align*}
so that a stationary point $\ell$ of $-\tilde{\Phi}-\tilde{S}$ must satisfy
$cc_p = -\log(1-\ell)/Ei\left(\frac{L(\ell)}{\hat{b}_p}\right)$. Since
\begin{align*}
\frac{d}{d\lambda}\frac{-\log(1-\lambda)}{Ei\left(\frac{L(\lambda)}{\hat{b}_p}\right)}
&=\frac{\frac{Ei\left(\frac{L(\ell)}{\hat{b}_p}\right)}{1-\lambda}
	-\frac{e^{-L(\lambda)/\hat{b}_p}\log(1-\lambda)}{\lambda\log(\lambda)}}
	{Ei\left(\frac{L(\ell)}{\hat{b}_p}\right)^2}\\
&\leq e^{-L(\lambda)/\hat{b}_p}
	\frac{
	\frac{\log\left(1-\frac{\underline{b}_p}
	{(\hat{b}_p-\underline{b}_p)\log(\lambda)}\right)}{1-\lambda}
	-\frac{\log(1-\lambda)}{\lambda\log(\lambda)}}
	{Ei\left(\frac{L(\ell)}{\hat{b}_p}\right)^2},
\end{align*}
the function $-\tilde{\Phi}-\tilde{S}$ admits at most one stationary point in $(0,1)$ if 
\begin{align}\label{ConditionI}
\frac{\log\left(1-\frac{\underline{b}_p}
	{(\hat{b}_p-\underline{b}_p)\log(\lambda)}\right)}{1-\lambda}
	-\frac{\log(1-\lambda)}{\lambda\log(\lambda)}< 0,
\Leftrightarrow
\log(\lambda)\left[1-(1-\lambda)^{\frac{1-\lambda}{\lambda\log(\lambda)}}\right]
  < \frac{\underline{b}_p}
	{(\hat{b}_p-\underline{b}_p)}.
\end{align}
Since, for $0<\lambda<1$, $\log(\lambda)\left[1-(1-\lambda)^{\frac{1-\lambda}{\lambda\log(\lambda)}}\right]<1.2$, condition \ref{ConditionI} holds if $0.55\hat{b}_p <\underline{b}_p$. Since
\begin{align*}
\lim_{\lambda\rightarrow 0^+} -\tilde{\Phi}(\lambda)
  & = \lim_{\lambda\rightarrow 0^+} \tilde{S}(\lambda)
	= 0,\ 
\lim_{\lambda\rightarrow 0^+} \frac{-\tilde{\Phi}(\lambda)}{\tilde{S}(\lambda)}
    = \infty\\
\lim_{\lambda\rightarrow 1^-} -\tilde{\Phi}'(\lambda)
  & = \lim_{\lambda\rightarrow 1^-} \tilde{S}'(\lambda)
	= \infty, \
\lim_{\lambda\rightarrow 1^-} \frac{-\tilde{\Phi}'(\lambda)}{\tilde{S}'(\lambda)}
    = 0,
\end{align*} 
it must be the case that if $\lim_{\lambda\rightarrow 1^-}-\tilde{\Phi}(\lambda)\geq \lim_{\lambda\rightarrow 1^-}\tilde{S}(\lambda)$, which is \ref{ConditionLower}, then $-\tilde{\Phi}-\tilde{S}$ admits a positive maximum in $(0,\lambda)$. Therefore, if $0.55\hat{b}_p <\underline{b}_p$ and \ref{ConditionLower} are satisfied, $-\tilde{\Phi}-\tilde{S}$ must be nonnegative on $(0,1)$, since it would otherwise admit two stationary points.
\end{proof}

We note that $0.55\hat{b}_p<\underline{b}_p$ for all the 16 representative points. The next two lemmas identify necessary and sufficient conditions for the case $b_p\geq \hat{b}_p$.

\begin{lemma}
Suppose $b_p\in[\hat{b}_p,\overline{b}_p]$. Then, it is necessary for $\nu_{b_p}\in\mathcal{N}$ to hold that 
\begin{align}
\gamma & > \frac{\overline{b}_p-\hat{b}_p}{\hat{b}_p}:=\tilde{\gamma} \label{ConditionUpper}\\
c & \leq \frac{1}{c_p} \label{ConditionUpper_c}.
\end{align}
\end{lemma}
\begin{proof}
Note that
\begin{align*}
\lim_{\lambda\rightarrow\infty}\Phi(\lambda)
=\lim_{\lambda\rightarrow\infty}S(\lambda)
=0,
\end{align*}
so, by l'Hopital's theorem, $\Phi(\lambda)\geq S(\lambda)$ implies 
\begin{align*}
\frac{S''(\lambda)}{\Phi''(\lambda)}=O(1)
\end{align*}
as $\lambda\rightarrow \infty$. Furthermore, using the implicit definition of $u$,
\begin{align*}
\Phi'(\lambda)
 & = \frac{1}{c}\log(u(\lambda)),\
\Phi''(\lambda)
   = \frac{u'(\lambda)}{cu(\lambda)}, \
S'(\lambda)
   = -c_pEi\left(\frac{L(\lambda)}{\hat{b}_p}\right),\
S''(\lambda)
   = c_p\frac{1}
	{\lambda^{\overline{b}_p/(\overline{b}_p-\hat{b}_p)+1}\log(\lambda)},
\end{align*}
and, using implicit differentiation,
\begin{align*}
u'(\lambda) 
& = \left(\frac{u^{2+1/\gamma}}{(1-u+\gamma)(1+\gamma)^{1/\gamma}}\right)
	\left(\frac{1}{(\lambda-1)}\right)^{2+1/\gamma}
  \sim \left(\frac{1}{(\gamma)(1+\gamma)^{1/\gamma}}\right)
	\left(\frac{1}{\lambda}\right)^{2+1/\gamma}.
\end{align*}
We thus need
\begin{align*}
2+\frac{1}{\gamma}<\frac{\overline{b}_p}{\overline{b}_p-\hat{b}_p}+1
\Rightarrow \gamma >\tilde{\gamma}.
\end{align*}
For \ref{ConditionUpper_c} simply note that 
\begin{align*}
\lim_{\lambda \rightarrow 1^+}\frac{\Phi(\lambda)}{S(\lambda)}=\frac{1}{cc_p}.
\end{align*}
\end{proof}

\begin{lemma}\label{ConditionUpperLemma} There is a function $\kappa_p:\left(\tilde{\gamma},\infty\right)\rightarrow (0,\tilde{c}]$, where
\begin{align*}
\tilde{c}:=\min\left\lbrace\lim_{\lambda\rightarrow 1^-}\frac{1}{\tilde{S}(\lambda)},\frac{1}{c_p}\right\rbrace,
\end{align*}
such that, for every $\gamma$ that satisfies \ref{ConditionUpper}, $\nu_{b_p}\in \mathcal{N}$ if $c<\kappa_p(\gamma)$ and $\nu_{b_p}\notin \mathcal{N}$ if $c>\kappa_p(\gamma)$.
\end{lemma}
\begin{proof} Fix $\gamma>\tilde{\gamma}$. As in the previous lemma, and since $u$ does not depend on $c$, there is $\ell>1$ independent of $c$ such that for every $\lambda>\ell$,
\begin{align*}
\frac{-(1-\lambda)\log(u(\lambda))
	+\left(1-u(\lambda)\right)^{1/(1+\gamma)}}
	{Ei\left(\frac{L(\lambda)}{\overline{b}_p}\right)
	- \lambda Ei\left(\frac{L(\lambda)}{\hat{b}_p}\right)}
	\geq 1.	
\end{align*}
Hence, if $c<\tilde{c}$, $\Phi(\lambda)>S(\lambda)$ for every $\lambda>\ell$. Since $\Phi-S$ is continuous and decreasing in $c$ for every $\lambda\in [1,\ell]$, with $\lim_{c\rightarrow 0} \Phi-S=\infty, \ \lim_{c\rightarrow \infty} \Phi-S=-S<0$, there is a bounded set of values $c>0$ such that $\Phi(\lambda)-S(\lambda)\geq 0$ for every $\lambda\in [1,\ell]$. Letting $\overline{c}$ denote the supremum of such values, one can set $\kappa(\gamma)=\min\{\overline{c},\tilde{c}\}$.
\end{proof}

\begin{figure*}
        \centering
        \begin{subfigure}[b]{0.5\textwidth}
            \centering
            \includegraphics[width=\textwidth]
            {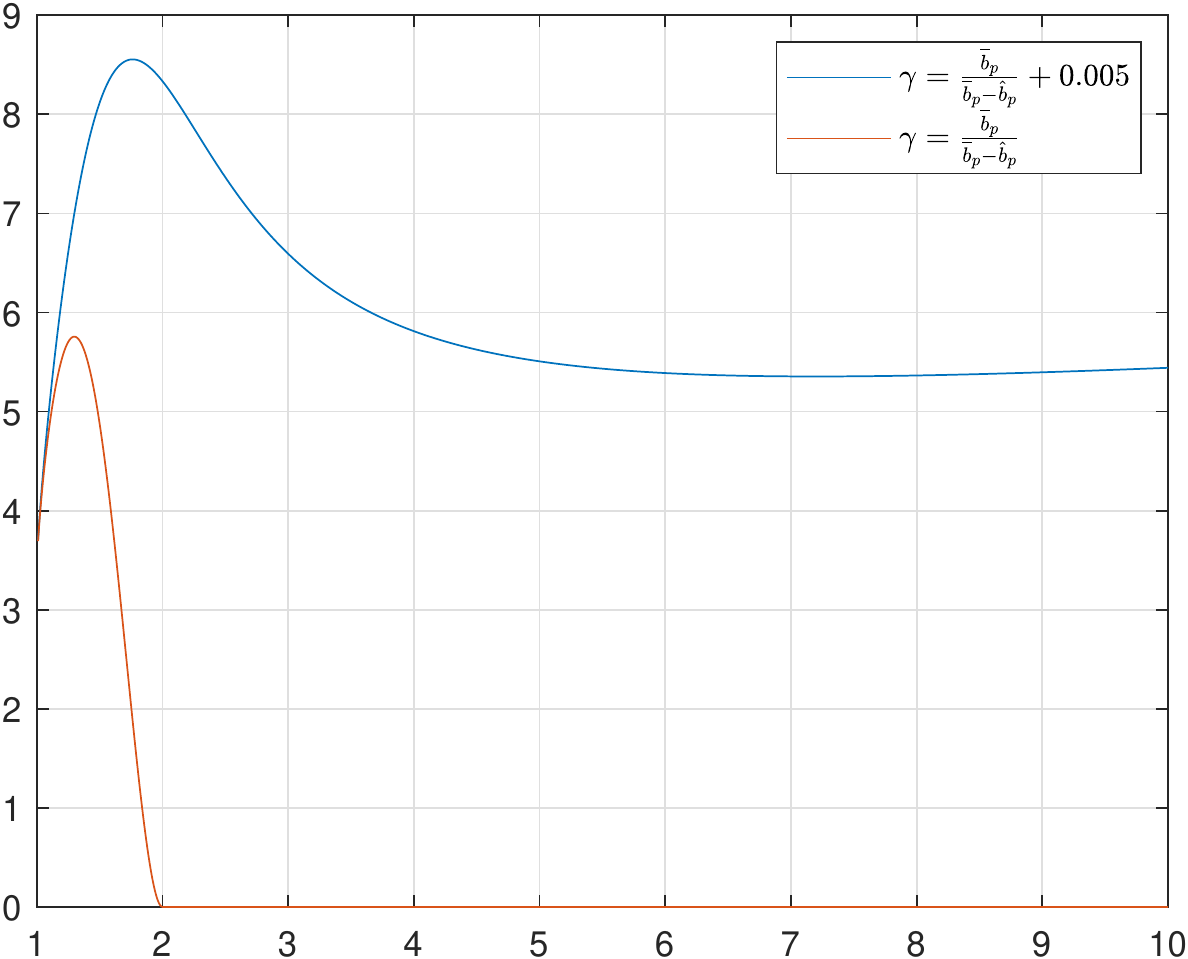}
            \caption[]%
            {{}}    
        \end{subfigure}
		\caption{The function $\Phi(\lambda)/S(\lambda)$ for the first of the 16 representative points, with $c=\tilde{c}$ and assuming $\gamma = \tilde{\gamma}+0.005$ (blue) and $\gamma=\tilde{\gamma}$ (red).}\label{cplot}
\end{figure*}

%\begin{algorithm}
%	\caption{Find $(c,\gamma)$ such that $\nu_{b_p}\in\mathcal{N}$ for every $b_p\in[\underline{b}_p,\overline{b}_p]$}\label{AlgoCGamma}
%	\begin{algorithmic}[1]
%		\Inputs {: $0<\alpha_c<1, \ \varepsilon_{\gamma}>0$}
%		\EndInput
%		\State {$\gamma = \tfrac{\overline{b}_p}{\overline{b}_p-\hat{b}_p}; \
%				c=\min\left\lbrace \tfrac{1}{c_p},\frac{1}{\tilde{S}(1)}\right\rbrace$}
%		\Do
%			\State {$c=\alpha_cc$}
%			\State {$\gamma=\gamma+\varepsilon_{\gamma}$}
%		\doWhile {$\min_{\lambda>1} \Phi(\lambda)-S(\lambda)<0$}
%	\end{algorithmic} 
%\end{algorithm}

The function $\kappa_p$ typically grows very fast, so that $\kappa_p(\gamma)=\tilde{c}$ for values of $\gamma$ that are slightly larger than $\tilde{\gamma}$. For instance, figure \ref{cplot} depicts the function $\Phi(\lambda/S(\lambda)$ for $\gamma = \tilde{\gamma}$ and $\gamma' = \tilde{\gamma}+0.005$, and for $c=\tilde{c}$ and the bilateral gamma parameters set as in the most representative of the quantized points. In fact, this is the case for each of the 16 quantized points, as shown in table \ref{table:CGamma}.

\begin{table}[h!]
\centering
 \begin{tabular}{| c | c | c || c | c | c || c | c | c || c | c | c |} 
 \hline $c$
 		& $\gamma$
 		& $\tilde{\gamma}$ 
 		& $c$
 		& $\gamma$
 		& $\tilde{\gamma}$
 		& $c$
 		& $\gamma$
 		& $\tilde{\gamma}$ 
 		& $c$
 		& $\gamma$
 		& $\tilde{\gamma}$\\
\hline\hline
0.462 & 0.357 & 0.347 & 0.212 & 0.259 & 0.219 & 0.075 & 0.400 & 0.380 & 0.138 & 0.288 & 0.258 \\ 
\hline
0.666 & 0.263 & 0.233 & 0.130 & 0.336 & 0.296 & 0.169 & 0.256 & 0.216 & 0.061 & 0.397 & 0.377 \\ 
\hline
0.262 & 0.252 & 0.232 & 0.126 & 0.327 & 0.287 & 0.069 & 0.484 & 0.424 & 0.039 & 0.387 & 0.357 \\ 
\hline
0.219 & 0.209 & 0.199 & 0.096 & 0.358 & 0.328 & 0.056 & 0.529 & 0.469 & 0.040 & 0.527 & 0.467 \\ 
\hline
\end{tabular}
\caption{Triples $(\tilde{c},\gamma,\tilde{\gamma})$ where $\gamma$ is the minimal value ensuring \ref{AcceptabilityConditionFrequency} with $c=\tilde{c}$.}
\label{table:CGamma}
\end{table}
 
\begin{rem}
Recalling that $\gamma$ is similar to the acceptability index for probability distortions, while $\tfrac{10}{c}$ roughly corresponds to the maximum distorted frequencies (so higher $c$ corresponds to smaller $\mathcal{N}$), we note that the values reported in \ref{table:CGamma} have the same magnitude and are consistent in general with those typically seen in the literature (see for instance \cite{EMPY}, \cite{MadanHDMT} and \cite{MDV}). 

\noindent We observe, in particular, that
\begin{itemize}
[nolistsep,noitemsep]
\item[i.] the three most representative points (top left corner of table \ref{table:CGamma}), are consistent with the pair $(0.25,0.25)$ used in \cite{MDV} to estimate capital requirements (chapter 15.5.2), and that the relatively high values of $\gamma$ is compensated by high values of $c$;
\item[ii.] for each triple, $\tilde{c}$ is higher, and in the less frequent cases close to, the value $0.01$, which, as shown in \cite{MadanHDMT}, generates higher returns compared to $c=1$ and $c=0.25$ for a portfolio constructed by maximization of the lower valuation.
\end{itemize}
\end{rem}

\section{Conclusions}
For an asset with (log) returns in the bilateral gamma class, a justification is provided, based on expected utility theory, that risks from holding the asset can be decomposed into a three dimensional vector of expected losses, variance of gains and variance of losses, while compensation for the risks is given by expected gains. Evidence is then provided that moments of bilateral gamma returns lie on a manifold with boundaries, and such boundaries are estimated via quantile and distorted linear and nonlinear regressions. It is observed that they imply a positive relationship between expected gains and variance of gains/expected losses, but a negative one between compensation and variance of losses thus implying market's operators being risk seekers in pure loss prospects. The claim that such finding are compatible with the experimental evidence that constitute prospects theory is then justified through a simple modification of Lucas Tree model. The analysis is corroborated by performing a similar one to the case of risk neutral parameters, assuming a separate drift to satisfy the martingale condition. An inverse relationship between shape and scale parameters of loss and gain process is observed and a theoretical boundary for scale parameters, in line with certain empirical observations, is described based on the theory of Conic finance. Finally, we observed that our estimates of the boundaries are generally larger than those implied by regulatory capital requirements.

\section{Acknowledgment}
This paper is a revised version of the second chapter of the author's doctoral dissertation, which was conducted under the supervision of Professor Dilip B. Madan at the Department of Mathematics of the University of Maryland, College Park.

\section{Appendix A: Assets Tickers}

The list of tickers of the assets considered in the empirical analyses performed in this research are reported in table \ref{table:Tickers} below.

\begin{table}[h!]
\centering
 \begin{tabular}{| c | c | c | c | c | c | c | c | c | c | c | c | c |} 
 \hline
 a & aapl & abc & abt & adbe & adm & aep & afl & akam & all & amat & amp & amt \\
 amzn & antm & aon & apa & apd & axp & ba & bac & bax & bby & bdx & ben & biib \\
 bk & bmy & c & cah & cat & ccl & cf & chrw & cl & cma  & cmcsa & cmi & cms \\
 cof & cop & cost & crm & csco & ctsh & ctxs & cvs & cvx & d & de & dgx & dhr \\
 dis & dov & duk & ebay & ecl & el & eog & eqt & etn & f & fcx &  fdx & fitb \\
 flr & fls & fslr & gd & ge & gild & gis & glw & gs & hal & hd & hes & hog \\
 hon & hp & hpq & hum & ibm & ice & intc & isrg & itw & ivz & jci & jnj & jnpr \\
 jpm & jwn & k & kim & klac & kmb & ko & kr & kss & lmt & lnc & low & m \\
 ma &  mcd &  mck & mdt & met & mmc & mmm & mo & mrk & mro & ms & msft & mtb \\
 mur & nem & nke & nov & nsc & ntap & nvda & nyt & orcl & oxy & payx & pcar & pfe \\
 pg & ph & pnc & ppg & pru & pxd & rf & rhi & rl & rok & rrc & sbux & schw \\
 slb & so & spg & spx & spy & stt & stz & syk & syy & t & tgt & tjx & tmo \\
 trv & txn & txt & unh & unp & ups & usb & vix & vlo & vno & vz & wfc & whr \\
 wmb & wmt & wy & x & xlb & xle & xlf & xli & xlk & xlp & xlu & xlv & xly \\
 xom & xrx & & & & & & & & & & & \\
 \hline
\end{tabular}
\caption{}
\label{table:Tickers}
\end{table}

\bibliographystyle{authordate1}
\bibliography{mybib_2011_pl,mybib_2011_pl_1,mybib_2011_pl_2,mybib2,mybib2_1,mybib2_2}

\end{document}